\documentclass[aps,prx,reprint,superscriptaddress,twocolumn,longbibliography,nofootinbib]{revtex4-2}

\usepackage[T1]{fontenc}
\usepackage{amsmath,amssymb,amsfonts,amsthm,mathtools}
\usepackage{microtype}
\usepackage{xcolor}
\usepackage[utf8]{inputenc}
\usepackage{comment}
\usepackage{graphicx}
\usepackage{tikz}
\usetikzlibrary{arrows.meta,positioning}

\usepackage[colorlinks=true]{hyperref}
\usepackage{booktabs}

\definecolor{darkgreen}{RGB}{30,135,65}
\definecolor{darkblue}{RGB}{0,0,170}
\definecolor{darkred}{RGB}{185,0,0}
\hypersetup{
  bookmarksnumbered,
  pdfstartview={FitH},
  citecolor=darkgreen,
  linkcolor=darkred,
  urlcolor=darkblue,
  pdfpagemode={UseOutlines}
}

\newtheorem{theorem}{Theorem}
\newtheorem{proposition}{Proposition}
\newtheorem{lemma}{Lemma}
\newtheorem{corollary}{Corollary}
\theoremstyle{definition}
\newtheorem{definition}{Definition}

\newcommand{\Tr}{\operatorname{Tr}}

\newcommand{\diag}{\operatorname{diag}}

\newcommand{\floor}[1]{\left\lfloor #1 \right\rfloor}
\newcommand{\ket}[1]{|#1\rangle}
\newcommand{\bra}[1]{\langle #1|}
\newcommand{\proj}[1]{|#1\rangle\langle #1|}

\newcommand{\tr}{\operatorname{Tr}}
\newcommand{\rank}{\operatorname{rank}}
\newcommand{\seq}{\mathrm{seq}}
\newcommand{\cmp}{\mathrm{cmp}}
\newcommand{\fresh}{\mathrm{fresh}}
\newcommand{\marg}{\mathrm{marg}}
\newcommand{\iid}{\mathrm{iid}}

\newcommand{\EP}{\mathrm{EP}}
\newcommand{\NO}{\mathrm{NO}}
\newcommand{\Uop}{\mathrm{U}}

\begin{document}

\title{The memory cost of sequential quantum state preparation}

\author{Jian Li}
\affiliation{Center for Theoretical Physics, Polish Academy of Sciences, Warsaw, Poland}
\author{Patryk Lipka-Bartosik}
\affiliation{Center for Theoretical Physics, Polish Academy of Sciences, Warsaw, Poland}
\affiliation{Institute of Theoretical Physics, Jagiellonian University, 30-348 Krak\'ow, Poland}

\date{\today }

\begin{abstract}
How much memory does a quantum device need to prepare systems one at a time? We address this question through the lens of quantum resource theories, which formalize physical restrictions on how the memory may interact with the stream of systems it processes. By comparing sequential with collective preparation, we show that under general assumptions both have the same asymptotic memory cost per output. Still, sequential preparation can require an unboundedly larger memory than collective preparation, and we investigate how this gap depends on the physical restrictions on the allowed interactions. For energy-conserving unitaries, we determine both optimal memory dimensions needed for preparing pure states from an energy eigenstate, and find that the extra memory required sequentially either remains bounded or grows logarithmically with the number of outputs. Under noisy operations, we reveal transformations for which one auxiliary qubit suffices for preparing arbitrarily many independent outputs collectively, while no memory of fixed dimension can deliver them sequentially. Finally, for sequential protocols that repeat a fixed interaction, we show that a $d$-dimensional memory that succeeds for $d^2$ rounds will succeed forever, and that such a sequential preparation is equivalent to quantum catalysis.

\end{abstract}

\maketitle

\section{Introduction} 
\label{sec:introduction}
Quantum devices often operate sequentially, preparing and releasing one system at a time. Examples include sources emitting successive photons~\cite{Lindner2009,Schwartz2016,Cogan2023}, refrigerators producing streams of cold qubits~\cite{Mitchison2015}, and devices using a finite phase reference to prepare coherent states~\cite{Bartlett2006a,Aberg2014,Vaccaro2018}. The systems retained between different preparations then form the device's memory. Depending on the operations allowed between rounds, this memory may need to supply energy, coherence, or other resources, as well as carry information between rounds. This leads to a basic question: How large must the memory be to prepare a given number of outputs under these operational constraints?

The answer is governed by the interplay of two constraints: physical restrictions on the available operations, and the requirement that outputs be released one at a time. Each constraint has been studied extensively, but rarely in combination. Quantum resource theories focus on determining the cost of state transformations under physical restrictions~\cite{Chitambar2019}, such as energy conservation~\cite{Chiribella2017,Gour_2008, Marvian_2013}, noisy~\cite{Horodecki2003, GOUR20151} and thermal operations~\cite{Brandao2013,Horodecki2013,Brandao2015,Czartowski2023,Czartowski2024,Lostaglio2022,Hsieh2026}, local operations and classical communication~\cite{Nielsen1999, Bennett1996}, or stabilizer operations~\cite{Bravyi2005, Veitch_2014}. The costs of preparing quantum states in these scenarios are well-understood either for a single copy, or for asymptotically many copies transformed jointly, with all inputs released together at the end.  Resource theories thus address the first constraint, but not the second.

Sequential state preparation, in turn, is the starting point of the theory of matrix product states and quantum combs~\cite{Schn2005a,Schn2007,Cirac2021,Gutoski2007,Chiribella2008,Chiribella2009}. There, an ancillary system interacts with one input at a time, and its dimension limits the correlations that can be built up between the outputs~\cite{Li2022}. Since the operations are typically unrestricted, however, this area addresses only the second constraint. More specifically, for independent outputs, no memory is needed at all, because each output can be prepared directly. Physical restrictions on allowed operations change this completely: a memory can become necessary even for independent outputs, because it must supply what the restricted operations cannot produce.  Symmetric tensor networks capture part of this effect for conserved quantities~\cite{Singh2011}, but more general restrictions lie outside their scope. The limits of sequential preparation under physical constraints therefore remain largely unexplored.

\begin{figure*}[h!t]
  \centering
  \includegraphics{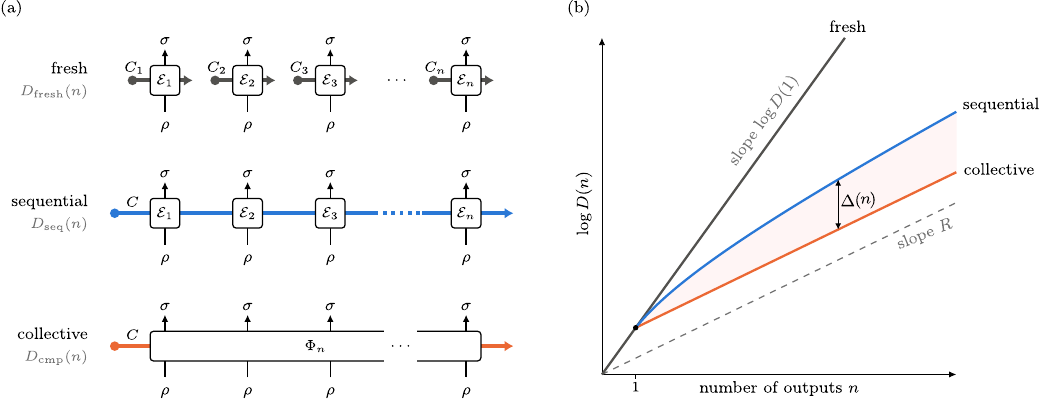}
  \caption{\textbf{Sequential preparation can need unboundedly more memory than collective preparation.} (a)~Three ways in which a memory $C$ (thick lines) assists the preparation of $n$ outputs in state $\sigma$ from inputs in state $\rho$, using allowed operations (boxes). In fresh preparation, each input has its own memory. In sequential preparation, a single memory meets the inputs one at a time, and each output is released before the next input arrives. In collective preparation, all inputs are processed jointly with the memory. (b)~Schematic memory costs $\log_2D(n)$, where $D(n)$ is the smallest dimension of a memory that suffices for $n$ outputs. The costs satisfy $D_{\mathrm{cmp}}(n)\le D_{\mathrm{seq}}(n)\le D_{\mathrm{fresh}}(n)=D(1)^n$ and coincide for a single output (dot). The sequential penalty $\Delta(n)$ is the gap between the sequential and collective costs (shaded). These two costs grow at the same rate $R$, the slope of the dashed line, so $\Delta(n)$ is sublinear in $n$, yet it can grow without bound.}
  \label{fig:setup}
\end{figure*}

Here we address this problem by comparing two ways of transforming $n$ inputs in state $\rho$ into $n$ outputs in state $\sigma$, using a memory and a given class of allowed operations (see Fig.~\ref{fig:setup}). In \emph{collective preparation} all inputs are processed jointly with the memory, as in standard resource theories. In \emph{sequential preparation}, the memory processes one input at a time and releases each output before the next input arrives. In both settings, the memory need not be restored: its state may change and become correlated with the outputs. We seek the minimal dimension of the memory that enables preparing $n$ outputs and denote the additional memory required by sequential preparation by \emph{sequential penalty}.

Our first result shows that sequential and collective preparation have the same asymptotic memory cost per output. This holds whenever the allowed operations are closed under composition and swapping systems is allowed. Consequently, the sequential penalty grows sublinearly with the number of outputs, but it can still grow without a bound. The cost of processing inputs sequentially is thus a finite-size effect that asymptotic rates cannot detect, and characterizing it requires accounting for how the state of the memory changes between rounds.

For interactions restricted to energy-conserving unitaries, we determine the minimum memory dimension for preparing arbitrary pure states from an energy eigenstate. For coherent targets, the memory can be understood as a finite phase reference (or a clock). We show that the energies occupied by the target determine which energy levels the memory must contain, and hence its minimum dimension. Sequential preparation then needs levels that enable releasing each output successively, whereas collective preparation can transfer energy to the entire batch at once.  This therefore fully characterizes the sequential penalty under energy-conserving unitaries. These results also determine the optimal lifetime of a $d$-dimensional reference: the maximum number of target states it can prepare sequentially.

When the system and memory interact through unitaries, supplemented with maximally-mixed ancillas (the setting of noisy operations), we show that the sequential penalty can diverge. More strikingly, we find transformations for which one auxiliary qubit suffices to prepare any sufficiently large batch of independent outputs collectively, yet no finite memory can produce them one at a time.

Finally, we connect sequential preparation to catalysis, which is the usual notion of reusability studied in quantum resource theories~\cite{Jonathan1999,LipkaBartosik2024,Datta2023,Son2024,Xing2024,deOliveiraJunior2025,obejko2024,Lie2021,Fang2026,Donnadieu2026,Son2024a}. We show that, for sequential protocols that repeat a fixed interaction, reusability and state restoration are equivalent. More specifically, a $d$-dimensional memory that succeeds for $d^2$ rounds is guaranteed to succeed forever, with a catalyst of the same dimension as the memory. Catalysis thus emerges as the stationary limit of sequential preparation.

The paper is organized as follows. Section~\ref{sec:model} introduces the framework, the preparation costs, and the sequential penalty. Section~\ref{sec:energy} determines the memory costs under energy conservation. Section~\ref{sec:finite} characterizes memory costs under noisy operations, and establishes the unbounded sequential penalty for certain families of states. Section~\ref{sec:rate} proves equality of the asymptotic rates, and Sec.~\ref{sec:stationary} develops the connection to catalysis.

\section{Sequential and collective preparation}\label{sec:model}

\subsection{Framework} \label{ssec:framework}
We fix an input state $\rho$ on system $A$, a target state $\sigma$ on system $B$, and a class of allowed operations, and compare three ways of preparing $n$ outputs with the assistance of a memory $C$ (Fig.~\ref{fig:setup}). In \emph{sequential preparation}, the memory interacts with one fresh input at a time and releases each output before the next input arrives. Released outputs cannot be accessed again, and all systems retained between rounds count as part of the memory. In \emph{collective preparation}, all $n$ inputs are processed jointly with the memory. As a baseline, we also consider \emph{fresh preparation}, where each input is processed with its own independently prepared memory. 

For each setting described above and number of outputs, the preparation cost is the minimum memory dimension, optimized over its initial state and all allowed operations. Notably, the memory need not be restored: its state may change, it may become correlated with the outputs, and its resources may be consumed.

Let $\Omega_{B_1\cdots B_n}$ denote the joint output state of the $n$-partite system, with reduced states $\Omega_{B_i}$. We distinguish two preparation requirements. \emph{Independent preparation} requires output states of the form
\begin{align}
  \Omega_{B_1\cdots B_n}=\sigma^{\otimes n},
\end{align}
whereas \emph{marginal preparation} requires only
\begin{align}
  \Omega_{B_i}=\sigma,\qquad i=1,\ldots,n,
\end{align}
allowing correlations between outputs. Fresh preparation automatically produces independent outputs, since each input is processed separately with an independently prepared memory. In sequential and collective preparation, even independent outputs may remain correlated with the memory. We use superscripts $\iid$ and $\marg$ to distinguish the two requirements when necessary. For a pure target, they coincide, and the outputs are uncorrelated with all remaining systems.

The physical restrictions are specified by a class $\mathcal O$ of allowed quantum channels, as in a quantum resource theory~\cite{Chitambar2019}. Throughout, we fix the transformation $\rho\to\sigma$ and the operation class $\mathcal O$, leaving this dependence implicit in our notation. The physical restrictions determine what the memory must supply and how it can change during preparation. To investigate how these restrictions affect the cost of sequential preparation, we will consider three classes of allowed operations: energy-preserving unitaries (Sec.~\ref{sec:energy}), noisy operations (Sec.~\ref{sec:finite}), and unrestricted unitaries (Sec.~\ref{sec:rate}). Quantities specific to these classes carry the superscripts $\EP$, $\NO$, and $\Uop$ respectively.

Despite their different physical restrictions, all classes of operations satisfy two assumptions used throughout:
\begin{itemize}
  \item[(A1)] \emph{Composition.}
  $\mathcal O$ contains the identity and is closed under
  composition and tensor products.

  \item[(A2)] \emph{Storage.}
  $\mathcal O$ permits swapping the input with a physically
  identical register held in the memory.
\end{itemize}
All Hilbert spaces are finite dimensional, and logarithms are base $2$. A memory of dimension $d$ therefore has logarithmic size $\log_2 d$ bits. The input and output systems satisfy $B\simeq A$, with $b:=\dim A=\dim B\ge2$.

\subsection{Sequential and collective protocols}
We now define the two preparation costs. In each case, the memory and the operations are optimized for a fixed number of outputs $n$ and a chosen output requirement.

\begin{definition}[Sequential protocol]\label{def:protocol}
A sequential protocol consists of a memory $C$ in an initial
state $\omega_0$ and channels
\begin{align}
  \mathcal E_i:\;A_iC\longrightarrow B_iC,
  \qquad \mathcal E_i\in\mathcal O,
  \qquad i=1,\ldots,n.
\end{align}
Each input $A_i$ is prepared in $\rho$, independently of the memory and all earlier outputs. Each channel acts only on $A_iC$, and the output $B_i$ is then released.
The protocol succeeds if the released outputs satisfy the chosen requirement (marginal or independent). The \emph{sequential cost} $D_{\seq}(n)$ is the smallest dimension of $C$ for which a successful protocol exists.
\end{definition}

Although the memory may become correlated with earlier outputs, each fresh input enters in a product state with it. The reduced memory state therefore evolves as 
\begin{align}\label{eq:memory-recursion}
  \omega_i
  =\tr_{B_i}\mathcal E_i(\rho\otimes\omega_{i-1}),
\end{align}
and the marginal of the output released in round $i$ is
\begin{align}
  \Omega_{B_i}
  =\tr_C\mathcal E_i(\rho\otimes\omega_{i-1}).
\end{align}
Independent preparation additionally constrains the joint state of the released outputs. Since later rounds never act on earlier output, satisfying either requirement for all outputs also satisfies it after every round.

\begin{definition}[Collective protocol]\label{def:collective}
A collective protocol consists of a memory $C$ in an initial
state $\nu$ and a channel
\begin{align}
  \Phi_n:\;A_1\cdots A_nC
  \longrightarrow B_1\cdots B_nC,
  \qquad \Phi_n\in\mathcal O.
\end{align}
The protocol succeeds if the output state
\begin{align}
  \Omega_{B_1\cdots B_n}
  =\tr_C\Phi_n(\rho^{\otimes n}\otimes\nu)
\end{align}
satisfies the chosen requirement.
The \emph{collective cost} $D_{\cmp}(n)$ is the smallest
dimension of $C$ for which a successful protocol exists.
\end{definition}
The two costs are optimized separately, so the states of their optimal memories may differ. We set $D_{\seq}(0)=D_{\cmp}(0)=1$. For a single output, sequential and collective preparation coincide, as do the marginal and independent requirements, namely
\begin{align}
  D_{\seq}^{\marg}(1)
  =D_{\seq}^{\iid}(1)
  =D_{\cmp}^{\marg}(1)
  =D_{\cmp}^{\iid}(1).
\end{align}
Differences between these costs can therefore arise only when preparing more than one output.

\subsection{The sequential penalty}
Under assumption (A1), composing the first $k$ rounds of a sequential protocol gives a collective protocol for $k$ inputs using the same memory. For either output requirement (independent or marginal), this implies 
\begin{align}\label{eq:prefix}
  D_{\seq}(n)\ge\max_{1\le k\le n}D_{\cmp}(k).
\end{align}
The inequality can be strict, even after optimizing over all initial memory states and allowed operations. We quantify this gap by the \emph{sequential penalty}, namely
\begin{align}\label{eq:penalty}
  \Delta(n):=\log_2\frac{D_{\seq}(n)}{D_{\cmp}(n)}\ge0,
\end{align}
where both costs refer to the same output requirement. It measures the extra memory, in bits, required to release the outputs one at a time. Our central question is therefore:
\begin{center}
\itshape
How much more memory does sequential preparation require over collective preparation?
\end{center}

Table~\ref{tab:summary} summarizes our answers.
Under energy-preserving unitaries
(Sec.~\ref{sec:energy}), we compute both costs for pure target states. The penalty either remains bounded or grows logarithmically in $n$, depending on the energy spectrum. Under noisy operations (Sec.~\ref{sec:finite}), we exhibit transformations for which independent sequential preparation requires an ever larger memory, even though the collective cost remains bounded. Finally, under (A1) and (A2), we show that $\Delta(n)=o(n)$ for either output requirement
(Sec.~\ref{sec:rate}). Sequential and collective preparation therefore have the same asymptotic memory cost per output. We evaluate this common rate in the settings studied below. Despite this common rate, sequential delivery can still require substantially more memory for a finite number of outputs, as quantified by the sequential penalty.

\subsection{Shared versus fresh memories}\label{subsec:fresh}
The sequential penalty compares the extra memory required
to prepare quantum states relative to collective preparation. To put this cost in perspective, we compare it with preparing each output separately. This establishes whether sequential processing still saves memory despite the penalty. We call a protocol \emph{fresh} if its memory consists of registers $C_1\otimes\cdots\otimes C_n$, initially in a product state, and round $i$ acts only on $A_iC_i$. The memory cost then counts the total dimension of all these registers.

Each register must support one preparation and therefore has dimension at least $a:=D_{\seq}(1)$. Conversely, $n$ independent optimal one-round protocols attain
\begin{align}\label{eq:fresh}
  D_{\fresh}(n)=a^n.
\end{align}
This benchmark is the same for marginal and independent preparation, since independent registers always produce independent outputs.

A simple fresh protocol stores one target state in each register and swaps it with the arriving input. Under (A2), this gives $a\le b$ and hence
\begin{align}\label{eq:chain}
  D_{\cmp}(n)\le D_{\seq}(n)\le D_{\fresh}(n)\le b^n.
\end{align}
Every finite task is, therefore, possible with sufficient memory. Fresh preparation sets the cost of treating each input separately, while collective preparation allows all inputs to be processed together. The sequential cost lies between them. Its position shows both how much memory is saved by sharing and how much additional memory is required by sequential delivery.

The ability to share the memory between rounds can in principle allow one to prepare more outputs than in the fresh strategy. Since $D_{\fresh}(n)=a^n$, separate memories with total dimension $d$ can support at most
\begin{align}
  N_{\fresh}(d)=\lfloor\log_a d\rfloor,
  \qquad a>1.
\end{align}
The corresponding lifetime for a shared memory is
\begin{align}\label{eq:lifetime}
  N_d:=\sup\{n:\,D_{\seq}(n)\le d\}.
\end{align}
Thus $N_d\ge N_{\fresh}(d)$, and comparing the two lifetimes shows how sharing extends operation within the same memory budget. If $a=1$, no memory is needed and the transformation
can be repeated indefinitely. Their ratio defines the \emph{sharing advantage},
\begin{align}\label{eq:advantage}
  \mathcal A(d):=\frac{N_d}{N_{\fresh}(d)},
  \qquad d\ge a>1.
\end{align}
It quantifies how much sharing improves the memory lifetime at fixed dimension.

The optimization allows a different protocol for each number of outputs. An infinite lifetime therefore means that arbitrarily long preparations are possible within the budget, without necessarily specifying a single protocol that works forever. We examine such sustained operation in Sec.~\ref{sec:stationary}.

\begin{table}[t]
\centering
\renewcommand{\arraystretch}{1.25}
\setlength{\tabcolsep}{4pt}
\begin{tabular}{lccc}
\toprule
Setting & $D_{\cmp}(n)$ & $D_{\seq}(n)$ & $\Delta(n)$\\
\midrule
$\EP$, $E=\{0,\epsilon\}$
& $n+1$ & $n+1$ & $0$\\
$\EP$, $E=\{\epsilon_1,\epsilon_2\}$
& $n+2$ & $\binom{n+2}{2}$ & $\log_2 n+O(1)$\\
$\NO$, qutrit family
& $O(1)$ & $\to\infty$ & $\to\infty$\\
$\Uop$, cooling
& $2^{n\Delta S+o(n)}$
& $2^{n\Delta S+o(n)}$
& $o(n)$\\
\bottomrule
\end{tabular}
\caption{Sequential penalty in the settings studied in this work: energy-preserving unitaries ($\EP$, Sec.~\ref{sec:energy}), with $\epsilon_1/\epsilon_2$ irrational, purity under noisy operations ($\NO$, Sec.~\ref{sec:finite}), and cooling under general unitaries ($\Uop$, Sec.~\ref{sec:rate}). The cooling example concerns full-rank states with entropy change $\Delta S:=S(\rho)-S(\sigma)>0$ in bits and requires only correct output marginals. The noisy-operation example requires independent outputs. More generally, in every operation class satisfying (A1) and (A2), we have $\Delta(n)=o(n)$ (see Sec.~\ref{sec:rate}).}
\label{tab:summary}
\end{table}

\section{Coherence under energy conservation}\label{sec:energy}
We first consider pure-state preparation under strict
energy conservation, where both memory costs can be
computed exactly. This setting shows how releasing
outputs one at a time can require a strictly larger memory, even when the final outputs are identical.

\subsection{Setting}
Each input is a system $S$ with Hamiltonian $H_S$, prepared in an energy eigenstate $\ket0$ whose energy we take as zero, $H_S\ket0=0$. The target is an arbitrary pure state
\begin{align}\label{eq:psi}
  \ket\psi
  =\sum_{\epsilon\in E}
    \sqrt{p_\epsilon}\,\ket{\psi_\epsilon}, \qquad
  H_S\ket{\psi_\epsilon}
  =\epsilon\ket{\psi_\epsilon},
\end{align}
where $E$ is the set of distinct occupied energies, the states $\ket{\psi_\epsilon}$ are normalized, and $p_\epsilon>0$ with $\sum_{\epsilon\in E}p_\epsilon=1$.

The memory has a Hamiltonian $H_C$, which is fixed during a protocol and optimized together with its initial state. Sequential operations are unitaries commuting with $H_S+H_C$, while collective operations commute with $\sum_{i=1}^n H_{S_i}+H_C$. The memory therefore supplies or absorbs any energy exchanged during preparation.

When the target occupies more than one energy, the memory must
also supply coherence between energy eigenspaces. It then acts
as a finite quantum reference frame for time
translations~\cite{BartlettRudolphSpekkens2007}, as
energy-preserving interactions cannot produce this coherence
from an incoherent initial state~\cite{Marvian2016}.
Under such interactions, a finite-dimensional reference cannot
generate coherence in an initially incoherent system while
leaving its own marginal state unchanged, even if correlations
with the output are allowed~\cite{LostaglioMuller2019}.
The degradation of finite references under repeated use has
been studied for specified reference states and interactions,
see e.g. Ref.~\cite{Bartlett2006}.
Here we optimize both the reference state and the interactions
to find the smallest memory capable of a prescribed number of
exact preparations.

Since the target is pure, every successful output is
uncorrelated with the memory and all other outputs.
The marginal and independent requirements therefore
coincide, and we omit their superscripts throughout
this section.

\subsection{Exact costs}
To determine the memory costs, we first identify the
energies carried by the target outputs. For $m$ outputs, the possible total energies form the set
\begin{align}
  mE:=\{\epsilon_1+\cdots+\epsilon_m \quad \text{s.t.} \quad \epsilon_1,\ldots,\epsilon_m\in E\},
\end{align}
with $0E:=\{0\}$. An energy may appear repeatedly in a sum, but each distinct total is counted only once. The set
\begin{align}\label{eq:Gamma}
  \Gamma_n:=\bigcup_{m=0}^{n}mE
\end{align}
then collects the possible total energies of up to $n$ outputs. The following theorem expresses memory costs in terms of these sets, reducing the optimization over states, Hamiltonians, and interactions to counting
distinct energies.

\begin{theorem}[Exact memory costs]\label{thm:energy}
For the preparation of $\ket\psi$ in Eq.~\eqref{eq:psi}
from $\ket0$ under energy-preserving unitaries,
\begin{align}\label{eq:energycost}
  D^{\EP}_{\seq}(n)&=|\Gamma_n|,
  &
  D^{\EP}_{\cmp}(n)&=|\{0\}\cup nE|
\end{align}
for every $n\ge0$. 
\end{theorem}
In particular, both memory costs depend only on the occupied energies $E$, not on their probabilities $p_\epsilon$. This is a consequence of requiring exact preparation, i.e. every occupied energy must be accommodated, however small its probability.

\emph{Proof sketch.} We first derive the lower bounds. Let $X_i$ denote the set of energies occupied by the memory after $i$ rounds, so $X_0$ is its initial energy support.

Consider round $i$. Before the interaction, the memory
occupies energies $X_{i-1}$ and the fresh input has zero energy. The joint state therefore has support on energies $X_{i-1}$. After the interaction, the output is
in the pure state $\ket\psi$ and is consequently uncorrelated with the memory. Every combination of an occupied output energy and an occupied memory energy then occurs, giving the total-energy support 
\begin{align}
  X_i+E:=\{x+\epsilon \quad \text{s.t.} \quad x\in X_i,\ \epsilon\in E\}.
\end{align}
Energy conservation equates the supports before and after the interaction, so $X_{i-1}=X_i+E$. Working backwards from the final round gives $X_{n-1}=X_n+E$, then $X_{n-2}=X_n+2E$, and hence 
\begin{align}\label{eq:layers}
  X_i=X_n+(n-i)E,\qquad i=0,\ldots,n.
\end{align}

Choose any final occupied memory energy $G\in X_n$. Eq.~\eqref{eq:layers} implies that the memory Hamiltonian must have eigenvalues at every energy in $G+mE$, for $m=0,\ldots,n$. Their union is $G+\Gamma_n$, which contains $|\Gamma_n|$ distinct energies. Since eigenstates with different energies are orthogonal, the memory dimension must be at least $|\Gamma_n|$. 

For collective preparation, the same reasoning relates the initial and final memory supports, $Y_{\rm in}$ and $Y_{\rm out}$, i.e.
\begin{align}
  Y_{\rm in}=Y_{\rm out}+nE.
\end{align}
Choosing any $G\in Y_{\rm out}$ shows that the memory Hamiltonian must contain both $G$ and every energy in $G+nE$. This requires at least $|\{0\}\cup nE|$ dimensions.

To attain the sequential bound, give the memory one eigenstate $\ket F$ at each energy $F\in\Gamma_n$. Let $p_F^{(m)}$ be the probability that $\ket\psi^{\otimes m}$ has total energy $F$, and define 
\begin{align}
  \ket{\Xi_m}
  :=\sum_{F\in mE}\sqrt{p_F^{(m)}}\,\ket F.
\end{align}
The states $\ket0\ket{\Xi_m}$ and $\ket\psi\ket{\Xi_{m-1}}$ have the same total-energy distribution. Specifically, the first encodes the energy of $m$ targets in the memory, while the second divides it between one target and the remaining $m-1$. Their components within each energy eigenspace therefore have equal norms. A unitary within that eigenspace can map one component to the other. Combining these unitaries gives an energy-preserving unitary satisfying 
\begin{align}
  U_m\ket0\ket{\Xi_m}
  =\ket\psi\ket{\Xi_{m-1}}.
\end{align}
Starting from $\ket{\Xi_n}$ and applying $U_n,U_{n-1},\ldots,U_1$ releases $n$ targets and leaves the memory in $\ket{\Xi_0}=\ket0_C$. This attains the sequential lower bound. 

For collective preparation, choose a memory Hamiltonian with one eigenstate at each energy in $\{0\}\cup nE$ and prepare it in $\ket{\Xi_n}$. Initially, all inputs have zero energy, and the memory carries the total energy distribution of $n$ targets. The same distribution is obtained when all outputs are in $\ket\psi$ and the memory is in $\ket0_C$. This guarantees the existence of an energy-preserving unitary implementing
\begin{align}
  \ket0^{\otimes n}\ket{\Xi_n}
  \longmapsto
  \ket\psi^{\otimes n}\ket0_C.
\end{align}
All targets are prepared in one operation, so the memory
needs no additional eigenstates for intermediate rounds.
Its dimension is $|\{0\}\cup nE|$, attaining the collective
lower bound. Appendix~\ref{app:energy} gives the
constructions in detail.

\subsection{When does sequential preparation require more memory?}
Sequential preparation can require extra memory because releasing outputs one at a time requires the memory traverse intermediate energies. If these energies are absent from the optimal collective memory, additional eigenstates are needed. Three examples illustrate when this happens.

\emph{An energy eigenstate $\ket{\psi} = \ket{\epsilon}$.}
Suppose each output must be excited to energy $\epsilon>0$,
so that $E=\{\epsilon\}$. A collective memory needs only two
energy eigenstates, at $n\epsilon$ and $0$. The optimal collective protocol transfers all its energy to the $n$ inputs in one operation. A sequential memory instead transfers $\epsilon$ in each round, passing through the energies $n\epsilon,(n-1)\epsilon,\ldots$ and $0$. Hence, for $n\ge1$,
\begin{align}
  D_{\cmp}^{\EP}(n)=2,\qquad D_{\seq}^{\EP}(n)=n+1.
\end{align}
In the optimal sequential protocol, the memory's energy
acts as a counter, i.e. after $i$ rounds its state is $\ket{(n-i)\epsilon}$, recording how many outputs remain to be prepared.

\emph{A coherent qubit $\ket{\psi} = \ket{+}$.} Let $H_S=\epsilon\ket1\!\bra1$ and consider a pure target with both energies occupied, such as $\ket+=(\ket0+\ket1)/\sqrt2$. Then $E=\{0,\epsilon\}$ and $mE=\{0,\epsilon,\ldots,m\epsilon\}$. In the construction above, the memory state $\ket{\Xi_m}$ with $m$ outputs remaining is a superposition of eigenstates with energies $0,\epsilon,\ldots,m\epsilon$. A memory with energies $0,\epsilon,\ldots,n\epsilon$ can therefore represent every state $\ket{\Xi_n},\ket{\Xi_{n-1}},\ldots,\ket{\Xi_0}$ needed
during sequential preparation. These are exactly the $n+1$ energies required for collective preparation, so
\begin{align}\label{eq:qubitcost}
  D_{\seq}^{\EP}(n)=D_{\cmp}^{\EP}(n)=n+1.
\end{align}
A memory with $n+1$ equally spaced energy levels therefore suffices in both collective and sequential scenario.

\emph{Incommensurate energies.} Consider any pure target $\ket{\psi}$ occupying two energies $\epsilon_1$ and $\epsilon_2$ with irrational ratio. The possible total energies of $m$ targets are 
\begin{align}
  mE=\{k\epsilon_1+(m-k)\epsilon_2 \quad \text{s.t.} \quad k=0,\ldots,m\},
\end{align}
giving $m+1$ distinct energies. Unlike the example of a coherent qubit, no energy in $mE$ appears in $m'E$ for $m'\ne m$. Consequently, the memory states $\ket{\Xi_m}$ used at
different rounds must occupy different energy
eigenstates. Sequential preparation requires all of them, whereas collective preparation needs only the $n+1$ energies in $nE$ and the final energy $0$. Theorem~\ref{thm:energy} therefore gives us, for any $n\ge1$,
\begin{align}\label{eq:triangular}
  D_{\seq}^{\EP}(n)
  =\sum_{m=0}^{n}(m+1)=\binom{n+2}{2}, \quad
  D_{\cmp}^{\EP}(n)=n+2.
\end{align}
The optimal sequential memory is therefore larger than the collective one by a factor $(n+1)/2$, giving a penalty
$\Delta(n)=\log_2[(n+1)/2]$. Here the optimal memory state also identifies how many outputs remain, i.e. an energy $k\epsilon_1+l\epsilon_2$ occurs when $k+l$ targets remain to be prepared. 

These examples lead to the following question: Can the same
total energy arise from different numbers of targets? If not, the total energy determines how many targets remain, and memory states at different rounds must occupy different energy eigenstates. For a target with energy $\epsilon>0$, the total $m\epsilon$ uniquely determines the number $m$.
For two energies with irrational ratio, the total $k\epsilon_1+l\epsilon_2$ uniquely determines both $k$ and $l$, and hence their sum. For the coherent qubit, however, adding a target can contribute zero energy, so the same total can occur for different numbers of targets.

The general criterion is therefore whether two sums of energies from $E$ can be equal while containing different numbers of terms. To quantify this, let $r$ be the largest number of rationally independent energies in $E$, and let $s$ be the largest number of rationally independent energy gaps $\epsilon-\epsilon'$ with $\epsilon,\epsilon'\in E$. Every gap is a difference of two energies, and every energy is a gap plus a fixed $\epsilon_0\in E$. Hence either $r=s$ or $r=s+1$. Equal sums with different numbers of terms exist exactly when $r=s$, that is, when zero is a rational affine combination of elements of $E$. For example, in the three cases studied above,  $(r,s)$ equals $(1,0)$ for a single energy, $(1,1)$ for the coherent qubit, and $(2,1)$ for two incommensurate energies. These two numbers determine how the memory costs grow.

\begin{proposition}[Memory growth under EP]\label{prop:energypenalty}
For every finite pure target with $E\neq\{0\}$, we have
\begin{align}\label{eq:energygrowth}
  D_{\seq}^{\EP}(n)=\Theta(n^r),\qquad
  D_{\cmp}^{\EP}(n)=\Theta(n^s).
\end{align}
\end{proposition}
Consequently, the sequential penalty satisfies
\begin{align}\label{eq:penaltyclassification}
  \Delta_{\EP}(n)=(r-s)\log_2n+O(1).
\end{align}
Since $r-s\in\{0,1\}$, the sequential penalty either remains bounded or grows as $\log_2n$. The two exponents have a simple origin. For any fixed $\epsilon_0\in E$, the total energies in $nE$ differ from $n\epsilon_0$ by sums of $n$ energy gaps, so a collective memory only needs to resolve combinations of gaps. A sequential memory must also hold the energies of every smaller number of remaining targets. When $r=s+1$, the total energy determines how many targets carry it, so each number of remaining targets requires its own set of memory energy eigenstates, as in the examples above. Including all these sets makes the sequential memory larger than the collective one by a factor proportional to $n$. When $r=s$, the sets overlap, allowing the same eigenstates to be used at different rounds. For details, see Appendix~\ref{app:sumsets}.

The penalty is also bounded for every finite $n$, namely \begin{align}\label{eq:energyfinitepenalty}
  0\le\Delta_{\EP}(n)\le\log_2(n+1).
\end{align}
This follows from counting the energies required at each round. For $m\le n$, shifting every energy in $mE$ by $(n-m)\epsilon$, with any fixed $\epsilon\in E$, gives distinct energies in $nE$. Thus $|mE|\le|nE|$. The union of the $n+1$ sets $0E,\ldots,nE$ consequently contains at most $(n+1)|nE|$ energies. Since the collective memory already requires at least $|nE|$ eigenstates, sequential preparation needs at most a factor $n+1$ more, or $\log_2(n+1)$ additional bits.

Although sequential preparation can require more memory than collective preparation, it still saves substantially over preparing each output separately. By Proposition~\ref{prop:energypenalty}, we can write $D_{\seq}^{\EP}(n)=\Theta(n^r)$ with $r\ge1$, whereas
\begin{align}
  D_{\fresh}^{\EP}(n)=a^n,
\end{align}
where $a=D_{\seq}(1)=|\{0\}\cup E|>1$. Thus sequential preparation retains a polynomial memory cost, compared with the exponential cost of separate preparations.

At fixed memory dimension $d$, this gives $N_d=\Theta(d^{1/r})$ preparations instead of $N_{\fresh}(d)=\lfloor\log_a d\rfloor$. For the case of coherent qubit preparation, this comparison is particularly simple. Namely, a shared $d$-dimensional memory can prepare at most $d-1$ outputs, while separate memories can only prepare $\lfloor\log_2d\rfloor$ of them. Hence we have
\begin{align}\label{eq:qrfadvantage}
  N_d=d-1, \quad \text{with} \quad
  \mathcal A(d)
  =\frac{d-1}{\lfloor\log_2d\rfloor}
    \longrightarrow\infty.
\end{align}
More generally, $\mathcal A(d)=\Theta(d^{1/r}/\log d)$, so the sharing advantage grows here without bound.

Sequential preparation can thus use far less memory than separate preparations, yet more than collective preparation. In the next section we turn our attention to a different physical scenario: \emph{noisy operations} (NO). Under these operations, we will find transformations for which a fixed memory dimension suffices for arbitrarily many independent outputs prepared collectively, but no fixed dimension suffices when they are released one at a time.

\section{Purity under noisy operations}\label{sec:finite}

We now study memory costs under a different set of physical constraints, as captured by \emph{noisy operations}~\cite{Gour2015}. In this setting one can append arbitrary systems in maximally-mixed states, apply arbitrary unitaries, and discard subsystems. These operations cannot increase purity and always preserve the maximally mixed state.

This setting reveals how sequential preparation depends both on the memory state after each round and on the correlations permitted between outputs. More specifically, we show that requiring outputs to be both independent and released one at a time can force the required memory dimension to grow without bound. We characterize exactly which transformations admit bounded sequential memory, and identify transformations for which a memory with fixed dimension suffices if the outputs are instead prepared collectively or allowed to be correlated.

\subsection{A minimal example of separation} \label{ssec:minimal}
A separation between preparation costs occurs already for two outputs. Consider diagonal states $\rho$ and $\sigma$ of a three-dimensional system, with spectra, respectively,
\begin{align}\label{eq:finitepq}
  p&=\Big(\tfrac25,\tfrac3{10},\tfrac3{10}\Big),
  &
  q&=\Big(\tfrac12,\tfrac25,\tfrac1{10}\Big).
\end{align}
For noisy operations, state conversion is governed by majorization: the sum of the largest $k$ input eigenvalues must be at least the corresponding output sum for every $k$~\cite{Gour2015}. Applying this criterion to the
joint system and memory gives the exact costs
\begin{align}\label{eq:finitegap}
  D_{\cmp}^{\NO}(2)=2,\qquad
  D_{\seq}^{\NO}(2)=3,\qquad
  D_{\fresh}^{\NO}(2)=4,
\end{align}
for both marginal and independent preparation. Appendix~\ref{app:finite} derives the lower bounds and constructs protocols attaining them. Thus, sharing a memory between rounds saves dimension relative to separate preparations, but sequential delivery still incurs a penalty $\Delta(2)=\log_2(3/2)$.

The reason a two-dimensional sequential memory fails is that the two rounds place incompatible demands on its intermediate state. Let $x$ be the largest eigenvalue of the memory state after the first round. The majorization bounds derived in Appendix~\ref{app:finite} imply
\begin{align}
  \text{successful first round}\implies &x\le\tfrac45,
  \\
  \text{successful second round}\implies &x\ge\tfrac56.
\end{align}
No memory state left by the first round can therefore enable the transformation in the second, even with arbitrary initial states, round-dependent operations, and correlations with the first output. On the other hand, collective preparation avoids this constraint and succeeds with a two-dimensional memory.

\subsection{Resource budgets}
A finite memory can support preparing quantum states by providing the resources necessary for transforming the inputs into outputs. Whenever each output requires a net increase in a resource, the memory's initial supply limits how many outputs can be prepared. The following principle formulates this intuition for both sequential and collective preparation. 

Let $F$ be additive on product states and nonincreasing
under the allowed operations. Assume also that the
memory can be freely replaced by a reference state
$\gamma$ with $F(\gamma)=0$. We obtain the following bound.
\begin{proposition}[Budget principle]\label{prop:budget}
If an allowed operation acting on
$\rho^{\otimes m}\otimes\omega$ produces $\sigma^{\otimes m}$, then
\begin{align}\label{eq:Fbudget}
  m\,[F(\sigma)-F(\rho)]\le F(\omega).
\end{align}
The outputs may remain correlated with the memory.
\end{proposition}

To see why the bound holds, replace the final memory by
$\gamma$, removing its correlations with the outputs.
The resulting allowed transformation is $\rho^{\otimes m}\otimes\omega \longmapsto \sigma^{\otimes m}\otimes\gamma.$ Additivity, monotonicity, and $F(\gamma)=0$ give Eq.~\eqref{eq:Fbudget}. For a sequential protocol, the same argument applies to any $m$ consecutive rounds.

When $F(\sigma)>F(\rho)$, each independent output requires a fixed resource increase. Hence, a finite memory budget $F(\omega)$ limits the number of outputs that can be prepared. For correlated outputs, individual resource costs need not add up, so the bound can be violated.

\subsection{Unbounded sequential cost}
For noisy operations, we can use a budget specified by
the smallest eigenvalue, namely
\begin{align}\label{eq:F}
  F(\tau):=-\log_2\!\bigl(r\lambda_{\min}(\tau)\bigr)
  =D_{\max}(u_r\|\tau),
\end{align}
where $\tau$ acts on an $r$-dimensional space and
$u_r$ is maximally mixed~\cite{Datta2009}. This function is additive and vanishes on $u_r$. It also cannot increase under any unital channel $\Phi$ between spaces of equal dimension, since 
\begin{align}
  \tau\ge\lambda_{\min}(\tau)I
  \quad\Longrightarrow\quad
  \Phi(\tau)\ge\lambda_{\min}(\tau)I.
\end{align}
If $\lambda_{\min}(\sigma)<\lambda_{\min}(\rho)$, the
resource cost per independent output is
\begin{align}
F(\sigma)-F(\rho)
=\log_2\frac{\lambda_{\min}(\rho)}{\lambda_{\min}(\sigma)}>0.
\end{align}
For a full-rank memory, Eq.~\eqref{eq:Fbudget} directly limits
its lifetime. A memory with zero eigenvalues evades this bound, but only while its occupied support can expand. With full-rank inputs, the memory rank cannot decrease. Whenever that rank stays constant, the round restricts to a unital channel on the occupied subspaces, where the same budget is finite. A finite memory can increase its rank only finitely often, so it cannot
produce independent outputs forever. 

\begin{corollary}[Unbounded sequential memory]\label{thm:support}
Let $\rho$ and $\sigma$ be full-rank states of the same
dimension with $\lambda_{\min}(\sigma)<\lambda_{\min}(\rho)$.
Under unital channels, the optimal lifetime $N_d$ for independent
preparation is finite for every finite memory dimension $d$.
Consequently,
\begin{align}
D_{\seq}^{\iid,\NO}(n)\longrightarrow\infty
\qquad\text{as }n\to\infty.
\end{align}
\end{corollary}
Decreasing the minimum eigenvalue therefore forces the 
sequential memory to grow. We next show that, for some such transformations, one memory qubit still suffices to prepare every sufficiently large batch collectively.

\subsection{Bounded collective cost}
We now consider transformations satisfying the same
condition $\lambda_{\min}(\sigma)<\lambda_{\min}(\rho)$.
Corollary~\ref{thm:support} rules out a bounded memory
for independent sequential preparation. We will show that,
for some of these transformations, one memory qubit nevertheless suffices when all inputs are processed together. To see this, note that Proposition~\ref{prop:budget} constrains the collective cost through the monotones
\begin{align}
  F_\alpha(\tau)
  :=\log_2\dim\tau-H_\alpha(\tau),
  \qquad \alpha\ge0,
\end{align}
where $H_\alpha$ is the R\'enyi entropy of order $\alpha$, with $H_1$ understood as the von Neumann entropy. This function is additive, nonincreasing under noisy operations, and zero on the maximally mixed state. For input and target states of the same dimension, the budget principle gives 
\begin{align}\label{eq:collectivebudget}
  n\,[H_\alpha(\rho)-H_\alpha(\sigma)]
  \le F_\alpha(\omega)\le\log_2d,
\end{align}
where $\omega$ is the initial state of a
$d$-dimensional memory.
Thus a bounded collective cost requires
$H_\alpha(\rho)\le H_\alpha(\sigma)$ for every
$\alpha\ge0$.

The following theorem gives sufficient conditions under which a single memory qubit enables preparing every sufficiently large batch. Note that no fixed memory used sequentially can implement such transitions indefinitely.

\begin{theorem}[Bounded collective, unbounded sequential memory]
\label{thm:family}
Let $\rho$ and $\sigma$ be full-rank states of the
same dimension satisfying
\begin{itemize}
  \item[(i)]
  $H_\alpha(\rho)<H_\alpha(\sigma)$ for every
  $\alpha\in(0,\infty]$, and
  $\det\rho<\det\sigma$;
  \item[(ii)]
  $\lambda_{\min}(\rho)>\lambda_{\min}(\sigma)$.
\end{itemize}
There is a finite $n_0$ such that
\begin{align}\label{eq:growinggap}
  D^{\iid,\NO}_{\cmp}(n)&=2
  \qquad(n\ge n_0),\\
  D^{\iid,\NO}_{\seq}(n)&\longrightarrow\infty
  \qquad(n\to\infty).
\end{align}
The collective transformation furthermore leaves the memory maximally mixed and uncorrelated with the outputs, 
\begin{align}\label{eq:collectivefamily-main}
  \rho^{\otimes n}\otimes\ket0\!\bra0
  \longmapsto
  \sigma^{\otimes n}\otimes u_2,
  \qquad n\ge n_0.
\end{align}
\end{theorem}

The theorem is proven in Appendix~\ref{app:family}. The collective obstruction can be understood by considering preparation without a memory. Converting $\rho^{\otimes n}$ into $\sigma^{\otimes n}$ requires
\begin{align}\label{eq:lambda_cond}
  \lambda_{\min}(\rho)^n
  \le \lambda_{\min}(\sigma)^n,
\end{align}
because a unital operation can never decrease the smallest eigenvalue. Condition (ii) violates this inequality for every $n$. Although both probabilities become exponentially small as $n$ grows, their ordering still forbids exact preparation. A pure memory qubit introduces zero eigenvalues into the joint input, removing this obstruction for every batch size. For a fixed batch, a sufficiently pure qubit can also remove the minimum-eigenvalue obstruction, but the required purity approaches one as the batch grows. Condition (i) then ensures that the remaining majorization inequalities hold for sufficiently large batches, as proved in Appendix~\ref{app:family}. The proof relies on large-deviation estimates based on asymptotic majorization~\cite{Jensen2019}.

\emph{A qutrit example.}
Qutrits are the smallest physical systems for which the conditions of Theorem~\ref{thm:family} can coexist. For qubits, $H_\infty(\rho)<H_\infty(\sigma)$ contradicts condition (ii). For qutrits, the following criterion reduces the family of entropy inequalities to a determinant comparison.
\begin{corollary}[Qutrits]\label{cor:qutrit}
Let $\rho$ and $\sigma$ be full-rank qutrit states with decreasingly ordered spectra $p$ and $q$ satisfying
$p_1>q_1$, $q_2\ge p_2$, and $p_3>q_3$. If $\det\rho<\det\sigma$, then the conditions of Theorem~\ref{thm:family} hold.
\end{corollary}
For example, let us take 
\begin{align}\label{eq:family}
  \rho&=\diag(0.70,\,0.15,\,0.15),\\
  \sigma&=\diag(0.45,\,0.45,\,0.10).
\end{align}
These spectra satisfy the ordering assumptions and $\det\rho < \det\sigma$. One qubit of memory therefore suffices for collective preparation of every sufficiently large batch. 

The contrast is particularly clear in this example. Collective preparation must overcome the forbidden decrease from $(0.15)^n$ to $(0.1)^n$ in the smallest eigenvalue of the batch. Sequential preparation must instead accommodate the decrease from $0.15$ to $0.1$ at every release. Whenever the memory rank stays constant, each independent output consumes at least $c=\log_2(0.15/0.10)$ bits of its min-eigenvalue budget. This cost does not decrease as the sequence grows. Since the budget is finite at each fixed rank and the rank can increase only finitely many times, every finite memory eventually becomes insufficient.

For the family in Theorem~\ref{thm:family}, the memory cost can be bounded by processing inputs in blocks. Appendix~\ref{app:family} gives an upper bound on this sequential memory cost,
\begin{align}
\log_2D_{\seq}^{\iid,\NO}(n)
\le 2\sqrt{n\log_2 b}+O(1),
\label{eq:familyupper}
\end{align}
where $b$ is the input dimension. The required dimension therefore grows without bound, but its logarithm grows at most as $\sqrt n$.

\section{Asymptotic memory costs}\label{sec:rate}
The preceding examples show that sequential delivery can require substantially more memory than collective preparation. Does this difference persist in the memory cost per output? It turns out that, under (A1) and (A2), one can easily show that the two costs have the same asymptotic rate. The sequential penalty $\Delta(n)$ can therefore diverge while contributing zero to this rate. We then evaluate this rate in the two settings studied here.

\subsection{A common rate for sequential and collective preparation}
Throughout this subsection, both costs refer to the same output requirement (i.e. either independent or marginal). Two sequential protocols can run one after the other using separate memories, while two collective protocols can process separate batches. Under (A1), this means 
\begin{align}\label{eq:dseq-submultiplicative}
  D_{\seq}(n+m)
  &\le D_{\seq}(n)\,D_{\seq}(m),\\
  D_{\cmp}(n+m)
  &\le D_{\cmp}(n)\,D_{\cmp}(m).
\end{align}
The logarithms of both costs are therefore subadditive. By Fekete's lemma, the cost per output, $n^{-1}\log_2D(n)$, converges to its infimum in each case. The following theorem shows that these limits agree.

\begin{theorem}\label{thm:rates}
Under (A1) and (A2), for either output requirement,
\begin{align}\label{eq:rates}
  \lim_{n\to\infty}\frac{\log_2D_{\seq}(n)}{n}
  &=\lim_{n\to\infty}\frac{\log_2D_{\cmp}(n)}{n}
  =R,\\
  R&:=\inf_{\ell\ge1}
       \frac{\log_2D_{\cmp}(\ell)}{\ell}.
\end{align}
Consequently, $\Delta(n)=o(n)$ for any resource theory satisfying (A1) and (A2).
\end{theorem}
We emphasize that the common rate $R$ may be different for marginal and independent preparations.

\emph{Proof sketch.} The simple bound from Eq.~\eqref{eq:prefix} already implies that the sequential rate cannot be smaller than the collective one. For the reverse inequality, we can construct a sequential protocol that uses a collective one in blocks. For that, fix a block of size $\ell$. The initial state of the memory contains a buffer that initially holds $\ell$ targets $\sigma$, together with independent copies of the memory state required by an optimal collective protocol for $\ell$ inputs.

In each round, the proposed sequential protocol swaps the fresh input with the next stored target, which is allowed by (A2). This releases one output and stores one input. After $\ell$ rounds, the buffer contains $\ell$ inputs. The collective converter then prepares a new batch of targets in the buffer using the memory register. Repeating this procedure delivers one target output at a time. 

More specifically, a buffer of dimension $b^\ell$ and $\lceil n/\ell\rceil-1$ registers in the converter suffice for $n$ outputs, giving
\begin{align}
  \log_2D_{\seq}(n)
  \le{}&\ell\log_2b +\bigl(\lceil n/\ell\rceil-1\bigr)
       \log_2D_{\cmp}(\ell). 
\end{align}
For fixed $\ell$, the buffer contributes a vanishing memory cost per output as $n\to\infty$. Each converter register then supplies $\ell$ outputs at a cost of $\log_2D_{\cmp}(\ell)$ bits. Taking the infimum over $\ell$ leads to Eq.~\eqref{eq:rates}. For more details see Appendix~\ref{app:rates}.

This simple construction allows us to bound the sequential penalty for any finite number of outputs. Define the excess cost of a finite block above the asymptotic rate by
\begin{align}
  g(\ell):=\log_2D_{\cmp}(\ell)-\ell R\ge0.
\end{align}
Since $\log_2D_{\cmp}(n)\ge nR$, we obtain
\begin{align}\label{eq:blockpenalty-main}
  \Delta(n)\le
  \min_{1\le\ell\le n}
  \left\{\ell\log_2b+\frac{n}{\ell}g(\ell)\right\}.
\end{align}
The first term is the buffer cost, while the second is the excess cost of the collective blocks. Larger blocks require a larger buffer but fewer converter registers.

If $g(\ell)$ remains bounded, choosing $\ell$ proportional to $\sqrt n$ gives $\Delta(n)=O(\sqrt n)$. This recovers the bound for the noisy-operation family from Eq.~\eqref{eq:familyupper}. This construction gives an upper bound on the sequential memory cost, which need not be optimal.

Equality of rates leaves open the question of whether sequential preparation requires no additional memory, a bounded overhead, or an overhead that grows with $n$. The sequential penalty distinguishes these possibilities, as illustrated in Secs.~\ref{sec:energy} and~\ref{sec:finite}.

\subsection{Vanishing and extensive memory costs.}
Theorem~\ref{thm:rates} reduces the leading sequential cost to a collective problem, which can be solved with standard tools. We give two examples at opposite extremes.

\emph{Purity under noisy operations.} For full-rank states of the same dimension, the common rate for independent outputs is given by~\cite{Jensen2019}
\begin{align}\label{eq:noisyrate}
R_{\seq}^{\iid,\NO}
=R_{\cmp}^{\iid,\NO}
=\sup_{\alpha\in[0,\infty]}
\big[H_\alpha(\rho)-H_\alpha(\sigma)\big]_+,
\end{align}
where $[x]_+=\max\{x,0\}$ (see Appendix~\ref{app:noisy} for details). When this rate is positive, a memory of $M$ qubits enables preparing $M/R_{\seq}^{\iid,\NO}+o(M)$ independent outputs. For the family of states used in Theorem~\ref{thm:family}, all positive-order R\'enyi entropies increase, so the rate vanishes. Nevertheless, the fact that the minimum eigenvalue decreases implies that the sequential memory cost must grow without bound. The constraint is associated with R\'enyi order $-\infty$, which is outside of the range of $\alpha$ in Eq.~\eqref{eq:noisyrate}. Thus, the total memory cost diverges even though its asymptotic cost per output is zero.

\emph{Unitary cooling.} Now let every round be an arbitrary unitary on the input and the memory, and require only correct output marginals. For full-rank states with
$\Delta S:=S(\rho)-S(\sigma)>0$, the common rate is
\begin{align}\label{eq:coolrate}
R_{\seq}^{\marg,\Uop}
=R_{\cmp}^{\marg,\Uop}
=\Delta S.
\end{align}
The lower bound follows from properties of entropy, i.e., each round the system's entropy is reduced by $\Delta S$, so the memory's entropy must increase by at least that amount. A memory of $M$ qubits can therefore prepare at most $M/\Delta S$ outputs. This is the entropy-compression bound that limits closed-system algorithmic cooling~\cite{Schulman1999,Boykin2002,Rybr2008,Li2026}.  For full-rank states, this bound is attainable to leading order (see Appendix~\ref{app:cooling} for details). The required memory size in bits consequently grows linearly with the number of outputs, while its sequential penalty remains sublinear by Theorem~\ref{thm:rates}. Unlike the noisy-operation result above, this rate permits correlations between outputs. We note that our construction does not establish the rate for independent preparation. Note that this setting is complementary to the standard cooling scenario, in which the machine starts in a thermal state and one asks how cold a single target can be made~\cite{Clivaz2019,Xuereb2025,Xuereb2023,Biswas2026,Allahverdyan2011}

The common rate also determines the benefit of sharing a memory rather than providing a fresh one for each output. Let $D_{\seq}(1)>1$ be the minimum memory dimension for one output (see Sec.~\ref{subsec:fresh}). Fresh memories require $n\log_2 D_{\seq}(1)$ bits for $n$ outputs, whereas a shared memory requires $nR+o(n)$ bits. Consequently, for $R>0$, the sharing advantage in Eq.~\eqref{eq:advantage} satisfies
\begin{align}\label{eq:advantagerate}
  \lim_{d\to\infty}\mathcal A(d)
  =\frac{\log_2 D_{\seq}(1) }{R}.
\end{align}
For $R=0$, the shared memory cost is sublinear in $n$, and this advantage diverges (Appendix~\ref{app:rates}). This, however, does not imply an infinite lifetime at fixed dimension. Sharing the memory becomes more beneficial as the available memory dimension grows. For example, suppose $D_{\seq}(n)\sim n^2$. A memory consisting of $n$ registers of dimension $b$ then supports approximately $b^{n/2}$ outputs
when shared, compared with only $n\log_2b/\log_2D_{\seq}(1)$ outputs using fresh memories. The advantage grows without bound as $n$ increases, although every fixed memory still has a finite lifetime.

\section{Catalysis as stationary operation}\label{sec:stationary}
Catalysis guarantees reuse by restoring an auxiliary state after a transformation. Our framework instead allows the memory to change, requiring only that it continue to produce the desired outputs. We now show that these two notions of reuse can be connected. More specifically, we show that if a finite memory applies the same interaction in every round and produces the correct output marginals for at least $d^2$ rounds, then there exist a correlated catalyst of the same dimension that enables this transformation indefinitely. 

\subsection{Stationary protocols}

The protocols of Definition~\ref{def:protocol} allow the interaction to change each round. We now restrict to a smaller class of protocols that apply the same channel repeatedly.

\begin{definition}[Stationary protocol]
\label{def:stationary}
A sequential protocol is \emph{stationary} if
$\mathcal E_i=\mathcal E$ for every round.
Its memory and output channels are
\begin{align}\label{eq:reduced-channels}
  \Lambda(X)&:=\tr_B\mathcal E(\rho\otimes X),\\
  \Sigma(X)&:=\tr_C\mathcal E(\rho\otimes X).
\end{align}
\end{definition}
Stationarity refers here to the interaction, not to the memory state. Starting from $\omega_0$, the memory evolves as $\omega_i=\Lambda(\omega_{i-1})$, and round $i+1$ produces output $\Sigma(\omega_i)$.

A memory state $\omega$ is called a \emph{correlated catalyst} if
\begin{align}\label{eq:corrcat}
  \Lambda(\omega)=\omega,\qquad \Sigma(\omega)=\sigma.
\end{align}
Its marginal is restored, although it may become correlated with the output \cite{Mueller2018, Shiraishi2021}. An \emph{ordinary (uncorrelated) catalyst} satisfies the stronger condition $\mathcal E(\rho\otimes \omega)=\sigma\otimes \omega$. The former guarantees correct output marginals, while the latter guarantees independent outputs decoupled from the memory.

\subsection{From stationary protocols to catalysis}
Suppose a stationary protocol produces the correct output in every round. The memory need not converge to a fixed state. Nevertheless, its average over rounds, 
\begin{align}
  \bar\omega_N:=\frac1N\sum_{j=0}^{N-1}\omega_j
\end{align}
still produces the correct output,
$\Sigma(\bar\omega_N)=\sigma$, and satisfies
\begin{align}
  \Lambda(\bar\omega_N)-\bar\omega_N
  =\frac{\omega_N-\omega_0}{N}.
\end{align}
As $N$ grows, the averaged memory state becomes arbitrarily close to being restored after one use, while still producing the required output marginal. Since the memory is finite, a subsequence of these averages always converges to some state $\omega$. By continuity, $\Lambda(\omega)=\omega$ and $\Sigma(\omega)=\sigma$, so $\omega$ is a correlated catalyst.

In fact, for a fixed interaction on a memory of dimension $d$, exact success in the first $d^2$ rounds already guarantees success in every subsequent round. 

\begin{theorem}[Stationary protocols and catalysis]
\label{thm:stationary}
Fix a stationary interaction $\mathcal E$ on a memory $C$ of dimension $d$, with channels $\Lambda$ and $\Sigma$ defined as in Eq.~\eqref{eq:reduced-channels}. The following are equivalent:
\begin{enumerate}
  \item[(i)] There exists an initial memory state $\omega_0$
  such that
  \begin{align}
    \Sigma\!\left(\Lambda^j(\omega_0)\right)=\sigma,
    \qquad j=0,\ldots,d^2-1.
  \end{align}
  
  \item[(ii)] There exists an initial memory state $\omega_0$ such that
  \begin{align}
    \Sigma\!\left(\Lambda^j(\omega_0)\right)=\sigma,
    \qquad j\ge0.
  \end{align}
  \item[(iii)] There exists a correlated catalyst
  $\omega$ on $C$, satisfying
  \begin{align}
    \Lambda(\omega)=\omega,\qquad \Sigma(\omega)=\sigma.
  \end{align}
\end{enumerate}
\end{theorem}
The proof of the theorem is given in Appendix~\ref{app:catalysis}. 

Theorem~\ref{thm:stationary} states that success in the first $d^2$ rounds of a sequential protocol guarantees an infinite lifetime, and hence also a corresponding correlated catalytic transformation. Moreover, if an initial memory state satisfies
\begin{align}
  \mathcal E\!\left(\rho\otimes\Lambda^j(\omega_0)\right)
  =\sigma\otimes\Lambda^{j+1}(\omega_0),
\end{align}
for $j=0,\ldots,d^2-1$, then similarily there exists an uncorrelated catalyst $\omega$ on $C$ such that
\begin{align}
  \mathcal E(\rho\otimes \omega)=\sigma\otimes \omega.
\end{align}
The qutrit example in Eq.~\eqref{eq:family}
also connects our memory separation to catalysis. In Sec.~\ref{sec:finite}, we required independent outputs and showed that one qubit suffices for collective preparation, but the sequential memory dimension must grow with the number of outputs. The collective protocol consumes the qubit's purity and is therefore not a catalytic protocol.

\section{Discussion and outlook}
\label{sec:discussion}

We have studied the memory required to prepare quantum systems
sequentially under restrictions on the allowed operations. The
memory may change after each use, become correlated with the
outputs, and consume its initial resources. For either independent
or marginal preparation, we showed that sequential and collective
protocols have the same asymptotic memory cost per output under
our composition and storage assumptions. Their memory dimensions
can nevertheless differ by an unbounded factor. The sequential
penalty captures this difference, while the sequential memory
cost also determines the optimal lifetime at a fixed memory
dimension.

The examples studied here exhibit distinct forms of memory
growth. For pure-state preparation from an energy eigenstate,
energy conservation determines both optimal memory dimensions
through the occupied target energies. The resulting sequential
penalty either remains bounded or grows logarithmically with
the number of outputs. Under noisy operations, we identified
transformations for which a single pure memory qubit suffices
for every sufficiently large collective batch, while sequential
preparation of independent outputs requires an unbounded memory
dimension. We also
characterized the independent-output rate under noisy operations
through R\'enyi entropies. For unitary cooling with full-rank
states and marginal outputs, the optimal rate instead equals
the entropy decrease per output.

A natural next step is to determine the sequential penalty more
precisely beyond the leading asymptotic rate. For the noisy-operation
family, our block construction gives
$\log_2 D_{\seq}(n)=O(\sqrt{n})$, while the converse establishes
only that $D_{\seq}(n)$ diverges. Matching lower bounds or improved
protocols would determine the actual growth of the required
memory. More generally, it would be useful to identify which
properties of the state transformation and the allowed operations
determine whether the sequential penalty vanishes, remains
bounded, or grows with the number of outputs.

Our connection to catalysis concerns sustained operation with a
fixed interaction. Theorem~\ref{thm:stationary} shows that exact
marginal preparation for the first $d^2$ rounds with a
$d$-dimensional memory guarantees indefinite operation and the
existence of a correlated catalyst of the same dimension.
The general framework also permits the interaction to change
between rounds. Extending the characterization to such protocols
would clarify when this additional control reduces the memory
requirements, and under which assumptions bounded memory implies
the existence of a finite-dimensional catalyst.

The role of correlations suggests another extension. Independent
preparation requires the joint output state to be a product,
whereas marginal preparation allows correlations between outputs. Related trade-offs between residual system--catalyst correlations
and catalyst resource requirements have been established in
correlated catalysis~\cite{RubboliTomamichel2022}.
An intermediate requirement could bound the mutual information
between each newly released system and the earlier outputs.
Determining whether arbitrarily weak allowed correlations can
change the asymptotic growth of the memory dimension would clarify
how the independence requirement limits the lifetime of a finite
memory.
\bibliographystyle{aps-fullnames}

\section{Acknowledgements} 
We thank Jian Wei Cheong, Dawei Lu, Andri Pradana and Yuhao Wang for helpful discussions and comments. This work is supported by the Polish National Science Centre through project Sonata 2023/51/D/ST2/02309. We used Claude Opus 5.5 (Anthropic) and ChatGPT Sol 6.1 (OpenAI) to assist with literature searches, language editing, checking and adapting proofs, and numerical exploration, including the search for minimal examples of the separations discussed in Sec.~\ref{ssec:minimal}. All proofs, references, and numerical results were independently verified by the authors, who take full responsibility for the content of this paper.

\bibliography{bibliography}

\appendix 

\section{Proof of Theorem~\ref{thm:energy}}\label{app:energy}
Here we prove Theorem~\ref{thm:energy} and explain the construction of the optimal memory state. We first construct protocols attaining the memory dimensions in Eq.~\eqref{eq:energycost}, and then show that no smaller memory can prepare the target stream of states. For that we use the setting of Sec.~\ref{sec:energy}, i.e. the input is an energy eigenstate $\ket0_S$ with zero energy, and the target is
\begin{align}
\ket\psi_S = \sum_{\epsilon\in E}\sqrt{p_\epsilon}\,
  \ket{\psi_\epsilon}_S.
\end{align}
Here $E$ is the set of occupied energies, $p_\epsilon>0$ are their probabilities, and $\ket{\psi_\epsilon}$ is a normalized vector in the eigenspace with energy $\epsilon$. Recall that $mE$ is the set of sums of $m$ elements of $E$, with $0E=\{0\}$, and $\Gamma_n=\bigcup_{m=0}^n mE$. 

\subsection{Two properties of pure-state preparation}
The proof uses two well-known properties of pure states. First, a pure output cannot be correlated with any other system. Second, two pure states can be connected by an energy-preserving unitary exactly when their total-energy distributions agree. The first property will allow us to derive the lower bounds by tracking occupied energies. The second will provide the constructions attaining them.

\begin{lemma}
\label{lem:pure-output}
Let $S^n=S_1\cdots S_n$. If every $S_i$ marginal of a state $\Omega_{RS^n}$ equals $\proj{\psi}$,
then
\begin{equation}
  \Omega_{RS^n}
  =\Omega_R\otimes\bigl(\proj{\psi}\bigr)^{\otimes n}.
  \label{eq:pure-output-factorization}
\end{equation}
\end{lemma}

\begin{proof}
Suppose first that a positive operator $\omega_{RX}$ has the pure marginal
$\omega_X=\proj{\psi}$. With $Q_X=I_X-\proj{\psi}$,
\begin{equation}
  \Tr[(I_R\otimes Q_X)\omega_{RX}]=0.
\end{equation}
Positivity implies that $\omega_{RX}$ has no support on
$R\otimes\operatorname{ran}Q_X$. Hence
$\omega_{RX}=\omega_R\otimes\proj{\psi}$. Applying the same argument
successively to $S_1,\ldots,S_n$ proves Eq.~\eqref{eq:pure-output-factorization}.
\end{proof}
In particular, exact pure outputs cannot remain correlated with the memory or with earlier outputs.

\begin{lemma}\label{lem:sector-equivalence}
Let $H_{\rm tot}=\sum_F F P_F$, where $P_F$ projects onto the
eigenspace of total energy $F$, and let $\ket{\phi}$ and
$\ket{\varphi}$ be normalized pure states. An energy-preserving unitary $V$ satisfying
$V\ket{\phi}=\ket{\varphi}$ exists if and only if
\begin{equation}
  \bra{\phi}P_F\ket{\phi}
  =\bra{\varphi}P_F\ket{\varphi}
  \quad\text{for every }F.
  \label{eq:sector-equivalence}
\end{equation}
\end{lemma}

\begin{proof}
An energy-preserving unitary acts independently in each total-energy sector, so equal sector probabilities are necessary.  This proves necessity.

Conversely, suppose the sector probabilities agree. Choose in each occupied sector a unitary mapping the normalized projection of $\ket{\phi}$ to that of $\ket{\varphi}$, and take their direct sum. To construct the unitary within an occupied sector, extend each of the two normalized projections to an orthonormal basis of that sector and map the first basis to the second. On sectors where both projections vanish, choose the identity. The direct sum of these sector unitaries commutes with $H_{\rm tot}$ and
maps $\ket{\phi}$ to $\ket{\varphi}$. This criterion is the specialization to time translations of pure-state equivalence under symmetry-preserving unitaries, see Ref.~\cite{MarvianSpekkens2014} for details.

The sequential protocol we describe releases one target per round through the energy-preserving unitary constructed in the
following lemma. Throughout, identity operators on the remaining tensor
factors are implicit in sums of Hamiltonians.
\end{proof}

\subsection{Energy distributions and memory states}
The total-energy probabilities of
$\ket{\psi}^{\otimes m}$ are
\begin{align}
  p_F^{(m)} := \sum_{\substack{\epsilon_1,\ldots,\epsilon_m\in E\\
  \epsilon_1+\cdots+\epsilon_m=F}}
  p_{\epsilon_1}\cdots p_{\epsilon_m},
  \qquad m\ge1.
  \label{eq:qrf-energy-distribution}
\end{align}
For $m=0$, set $p_0^{(0)}=1$ and $p_F^{(0)}=0$ for
$F\ne0$. Thus $p_F^{(m)}>0$ exactly when $F\in mE$.
Separating one target from the remaining $m-1$
targets gives
\begin{align}
  p_F^{(m)}
  =
  \sum_{\epsilon\in E}
  p_\epsilon p_{F-\epsilon}^{(m-1)},
  \qquad m\ge1.
  \label{eq:qrf-recursion}
\end{align}
Probabilities outside their support are understood
to be zero.

Whenever a memory $C$ contains an orthonormal energy
eigenstate $\ket F_C$ at every energy $F\in mE$, we define
\begin{equation}
  \ket{\Xi_m}_C
  :=
  \sum_{F\in mE}\sqrt{p_F^{(m)}}\,\ket F_C.
  \label{eq:app-memory-states}
\end{equation}
This is a normalized pure state with the same energy
distribution as $m$ target systems. It retains coherence
between distinct occupied energies, using only one
memory eigenstate for each total energy.
For $m=0$, it is the zero-energy memory state
$\ket{\Xi_0}_C=\ket0_C$.

The collective construction below uses $\ket{\Xi_n}_C$
and $\ket0_C$. The sequential construction also needs
the intermediate states
$\ket{\Xi_{n-1}}_C,\ldots,\ket{\Xi_1}_C$ which are defined analogously. The two memories are chosen separately, with spectra that accommodate the states required in each setting.

\subsection{Optimal constructions and lower bounds}
In this section we prove Theorem~\ref{thm:energy}. We first construct protocols attaining the memory dimensions in Eq.~\eqref{eq:energycost}, and then show that no smaller memory suffices. Throughout, identity operators on the remaining tensor factors are implicit in sums of Hamiltonians.

The following lemma will be useful in what follows. It gives a construction for an energy-preserving unitary that releases one target per round.

\begin{lemma}\label{lem:release}
Let $C$ be a memory with one orthonormal eigenstate $\ket F_C$ for each $F\in\Gamma_n$, and Hamiltonian
\begin{align}
  H_C=\sum_{F\in\Gamma_n}F\,\proj{F}_C.
  \label{eq:app-seq-memory}
\end{align}
For every $m=1,\ldots,n$, there is a unitary $U_m$ commuting with $H_S+H_C$ such that
\begin{align}
  U_m\bigl(\ket0_S\ket{\Xi_m}_C\bigr)
  =\ket\psi_S\ket{\Xi_{m-1}}_C.
  \label{eq:app-release-target}
\end{align}
\end{lemma}
\begin{proof}
Fix $m\in\{1,\ldots,n\}$. Since $mE$ and $(m-1)E$ are subsets of
$\Gamma_n$, the memory accommodates both $\ket{\Xi_m}_C$ and
$\ket{\Xi_{m-1}}_C$, defined in Eq.~\eqref{eq:app-memory-states}. The
components $\ket{\psi_\epsilon}_S$ of the target in Eq.~\eqref{eq:psi} are
normalized and mutually orthogonal since they belong to distinct eigenspaces of $H_S$.

We define a unitary $U_m$ by mapping the input, together with each memory
eigenstate $\ket F_C$ with $F\in mE$, onto the normalized component of $\ket\psi_S\ket{\Xi_{m-1}}_C$ with total energy $F$. More specifically, we have
\begin{align}
  U_m\bigl(\ket0_S\ket F_C\bigr)
  :=\sum_{\substack{\epsilon\in E\\ F-\epsilon\in(m-1)E}}
  &\sqrt{p_\epsilon}\,
  \sqrt{\frac{p_{F-\epsilon}^{(m-1)}}{p_F^{(m)}}}\\
  &\times\ket{\psi_\epsilon}_S\ket{F-\epsilon}_C .
\label{eq:app-release-unitary}
\end{align}
The condition under the sum selects the terms with
$p_{F-\epsilon}^{(m-1)}>0$, so every memory ket $\ket{F-\epsilon}_C$ is
one of the memory eigenstates. Terms with different $\epsilon$ are
orthogonal because their system factors are. The recursion in
Eq.~\eqref{eq:qrf-recursion} therefore gives
\begin{align}
  \bigl\|U_m\bigl(\ket0_S\ket F_C\bigr)\bigr\|^2
  =\sum_{\substack{\epsilon\in E\\ F-\epsilon\in(m-1)E}}
  \frac{p_\epsilon\,p_{F-\epsilon}^{(m-1)}}{p_F^{(m)}}
  =1.
  \label{eq:app-release-normalization}
\end{align}
The input $\ket0_S\ket F_C$ has total energy $F$, and so does every term
on the right of Eq.~\eqref{eq:app-release-unitary}, since
$\epsilon+(F-\epsilon)=F$. Each prescription thus maps a unit vector to a
unit vector within the same eigenspace of $H_S+H_C$, and distinct values
of $F$ correspond to orthogonal eigenspaces. As in the proof of
Lemma~\ref{lem:sector-equivalence}, we complete each such pair of vectors
to orthonormal bases of its eigenspace and map the first basis onto the
second. On all other eigenspaces, $U_m$ acts as the identity. The
resulting direct sum is a unitary satisfying
Eq.~\eqref{eq:app-release-unitary} and
\begin{align}
  [U_m,H_S+H_C]=0.
  \label{eq:app-release-energy}
\end{align}

It remains to apply $U_m$ to $\ket0_S\ket{\Xi_m}_C$. By linearity, the
amplitudes $\sqrt{p_F^{(m)}}$ of $\ket{\Xi_m}_C$ cancel the denominators
in Eq.~\eqref{eq:app-release-unitary}, giving
\begin{align}
  U_m\bigl(\ket0_S\ket{\Xi_m}_C\bigr)
  &=\sum_{F\in mE}\,
  \sum_{\substack{\epsilon\in E\\ F-\epsilon\in(m-1)E}}
  \sqrt{p_\epsilon\,p_{F-\epsilon}^{(m-1)}}
  \notag\\
  &\quad\times\ket{\psi_\epsilon}_S\ket{F-\epsilon}_C .
  \label{eq:app-release-action}
\end{align}
Under the substitution $G=F-\epsilon$, the pairs $(F,\epsilon)$ in this
double sum correspond one-to-one to the pairs
$(\epsilon,G)\in E\times(m-1)E$, since $G+\epsilon\in mE$ whenever
$\epsilon\in E$ and $G\in(m-1)E$. The right-hand side of
Eq.~\eqref{eq:app-release-action} therefore equals
\begin{align}
  &\sum_{\epsilon\in E}\,\sum_{G\in(m-1)E}
  \sqrt{p_\epsilon\,p_G^{(m-1)}}\,
  \ket{\psi_\epsilon}_S\ket G_C
  \notag\\
  &\hspace{2cm}
  =\ket\psi_S\ket{\Xi_{m-1}}_C,
  \label{eq:app-release-factorization}
\end{align}
where we used Eqs.~\eqref{eq:psi} and~\eqref{eq:app-memory-states}.
\end{proof}

We are now ready to prove Theorem~\ref{thm:energy}.
\begin{proof}
The proof is divided into three parts. \\
\emph{Part I: Collective achievability.} For collective preparation, the memory $C$ carries a Hamiltonian $H_C$, and the allowed unitaries $V_n$ on $CS_1\cdots S_n$ satisfy
\begin{align}
  \Bigl[V_n,\,H_C+\sum_{j=1}^nH_{S_j}\Bigr]=0.
  \label{eq:qrf-block-energy-preservation}
\end{align}
We choose a memory with one orthonormal eigenstate $\ket{F}_C$ for each
$F\in\{0\}\cup nE$, and Hamiltonian
\begin{align}
  H_C=\sum_{F\in\{0\}\cup nE}F\proj{F}_C,
\end{align}
and initialize it in $\ket{\Xi_n}_C$. The pure states
\begin{align}
  \ket{\Xi_n}_C\ket{0}_S^{\otimes n}
  \quad\text{and}\quad
  \ket{0}_C\ket{\psi}_S^{\otimes n}
  \label{eq:qrf-block-states}
\end{align}
have the same total-energy distribution $p_F^{(n)}$. In the first state, the total energy is carried entirely by the memory, which has energy $F$ with probability $p_F^{(n)}$. In the second state, the memory has zero energy, and the $n$ targets have total energy $F$ with probability $p_F^{(n)}$. Lemma~\ref{lem:sector-equivalence} therefore provides an energy-preserving unitary $V_n$ mapping the first state to the second. This memory has dimension $|\{0\}\cup nE|$, so
$D_{\cmp}^{\EP}(n)\le|\{0\}\cup nE|$.

\emph{Part II: Sequential achievability.} Take the memory of Lemma~\ref{lem:release}, initialized in $\ket{\Xi_n}_C$, and apply $U_{n-i+1}$ to the memory and the fresh input $S_i$ in round $i=1,\ldots,n$. By Eq.~\eqref{eq:app-release-target}, round $i$ maps $\ket{\Xi_{n-i+1}}_C\ket{0}_{S_i}$ to $\ket{\Xi_{n-i}}_C\ket{\psi}_{S_i}$, so the $n$ rounds implement
\begin{align}
  \ket{\Xi_n}_C\ket{0}_{S_1}\cdots\ket{0}_{S_n}
  \longmapsto
  \ket{0}_C\ket{\psi}_{S_1}\cdots\ket{\psi}_{S_n}.
  \label{eq:qrf-sequential-chain}
\end{align}
In each round, the memory changes from $\ket{\Xi_m}_C$ to $\ket{\Xi_{m-1}}_C$ while producing one fresh copy of $\ket{\psi}$, uncorrelated with the memory and with the earlier outputs. After $n$ rounds, it reaches $\ket{\Xi_0}_C=\ket{0}_C$, while all $n$ targets have been produced. Unlike in Part~I, the memory must also hold the intermediate states $\ket{\Xi_{n-1}}_C,\ldots,\ket{\Xi_1}_C$, which is why its spectrum is $\Gamma_n$. It has dimension $|\Gamma_n|$, so $D_{\seq}^{\EP}(n)\le|\Gamma_n|$.

\emph{Part III: Converse.}
Let $C$ be the $d$-dimensional memory of a successful collective protocol, or of a successful sequential protocol, for $n$ outputs. Write its Hamiltonian as $H_C=\sum_GGQ_G$, where $Q_G$ is the energy-$G$ spectral projector, and define the energy support of a memory state $\omega_C$ by
\begin{align}
  X(\omega_C):=\{G:\Tr(Q_G\omega_C)>0\}.
  \label{eq:qrf-final-auxiliary-support}
\end{align}
This set is nonempty and contained in the spectrum $\operatorname{spec}(H_C)$, which has at most $d$ distinct values. For the collective protocol, let $U=V_n$ act on $CS_1\cdots S_n$, and set $H_{\rm tot}:=H_C+\sum_{j=1}^nH_{S_j}$. For the sequential protocol, let $U$ be the unitary of a single round $i$, acting on $CS_i$, and set $H_{\rm tot}:=H_C+H_{S_i}$. In both cases, $[U,H_{\rm tot}]=0$. If $P_F$ projects onto the eigenspace of $H_{\rm tot}$ with energy $F$, then, for
every joint state $\Omega$, 
\begin{align}
  [U,P_F]=0,
  \qquad
  \Tr(P_FU\Omega U^\dagger)=\Tr(P_F\Omega)
  \quad\text{for every }F.
  \label{eq:qrf-sector-conservation}
\end{align}
Thus the total-energy probability distribution, and in particular its support, is exactly conserved. For a product state, this support consists of all sums of energies occupied by the factors. In the collective protocol, the fresh inputs have zero energy and the memory starts in a state $\nu$, so the initial total-energy support is $Y_{\rm in}:=X(\nu)$. By Lemma~\ref{lem:pure-output}, the final state is 
\begin{align}
  \nu'\otimes\bigl(\proj{\psi}\bigr)^{\otimes n},
\end{align}
where $\nu'$ is the final memory state. Its total-energy support is $Y_{\rm out}+nE$, with $Y_{\rm out}:=X(\nu')$. Exact conservation equates the two supports. For any $G_0\in Y_{\rm out}$, the spectrum of $H_C$ therefore contains both $G_0$ and $G_0+nE\subseteq Y_{\rm in}$, that is, 
\begin{align}
  G_0+\bigl(\{0\}\cup nE\bigr)\subseteq\operatorname{spec}(H_C).
  \label{eq:qrf-support-translate}
\end{align}
Since translations preserve cardinality, $d\ge|\{0\}\cup nE|$.

In the sequential protocol, let $\omega_i$ be the memory state after round $i$, and write $X_i:=X(\omega_i)$. The fresh input $S_i$ has zero energy and is uncorrelated with the memory, so before round $i$ the state of $CS_i$ is $\omega_{i-1}\otimes\proj{0}$, with total-energy support $X_{i-1}$. By Lemma~\ref{lem:pure-output}, the state after the round is $\omega_i\otimes\proj{\psi}$, with total-energy support $X_i+E$. Exact conservation gives $X_{i-1}=X_i+E$, and iterating backwards from the final round yields Eq.~\eqref{eq:layers}. Since the memory Hamiltonian is the same in every round, for any $G_0\in X_n$ its spectrum contains $G_0+mE\subseteq X_{n-m}$ for every $m=0,\ldots,n$, that is, 
\begin{align}
  G_0+\Gamma_n\subseteq\operatorname{spec}(H_C).
  \label{eq:qrf-sequential-support}
\end{align}
Hence $d\ge|\Gamma_n|$.

This proves both lower bounds in Eq.~\eqref{eq:energycost}. A collective memory must therefore distinguish all total energies that the $n$ outputs can carry, together with its final energy. A sequential memory must do so for every number $m\le n$ of outputs still to be prepared.

\end{proof}

\section{Proof of Proposition~\ref{prop:energypenalty}}
\label{app:sumsets}
Here we prove Proposition~\ref{prop:energypenalty}, together with the
criterion for equal sums and the lifetime scaling stated in
Sec.~\ref{sec:energy}. By Theorem~\ref{thm:energy}, both sequential and collective memory costs depend on the number of distinct total energies, i.e. $D_{\seq}^{\EP}(n)=|\Gamma_n|$ and $D_{\cmp}^{\EP}(n)=|\{0\}\cup nE|$. We estimate them here using a simple counting argument.

\emph{Counting argument.} Consider a finite set of energies $A$ and fix one element, say $a_0 \in A$. Every sum of $n$ elements $x_1,\ldots,x_n\in A$ can be written as
\begin{align}\label{eq:app-sum-differences}
  \sum_{t=1}^n x_t=na_0+\sum_{t=1}^n\left(x_t-a_0\right).
\end{align}
Let $a_1-a_0,\ldots,a_k-a_0$ be a maximal set of rationally independent differences so that all other differences are rational combinations of them. Rescaling by a common denominator, we can therefore write
\begin{align}\label{eq:app-integer}
  a-a_0=\sum_{j=1}^k q_j(a)\,\omega_j,
  \qquad q_j(a)\in\mathbb Z,
\end{align}
for every $a\in A$, where $\omega_1,\ldots,\omega_k$ are rationally independent. Inserting Eq.~\eqref{eq:app-integer} into Eq.~\eqref{eq:app-sum-differences} shows that a sum of $n$ elements of $A$ is specified by the $k$ integers $Q_j:=\sum_{t=1}^nq_j(x_t)$. Different sums correspond to different tuples $(Q_1,\ldots,Q_k)$, and each $Q_j$ has magnitude at most $hn$, where $h:=\max_{j,a}|q_j(a)|$. Hence there are at most $(2hn+1)^k$ distinct sums. Conversely, taking $j_i$ copies of each $a_i$ and $n-\sum_ij_i$ copies of $a_0$, with integers $j_i\ge0$ and $\sum_ij_i\le n$, gives $\binom{n+k}{k}\ge n^k/k!$
distinct sums. Consequently, the number of distinct sums of $n$ elements of $A$ scales as $\Theta(n^k)$. 

\begin{proof}[Proof of Proposition~\ref{prop:energypenalty}]
We now apply the above counting argument to compute the sequential and collective memory costs. For the sequential cost, observe that $\Gamma_n$ consists of all sums of $n$ elements of the set $A = \{0\} \cup E$. Choosing $a_0 = 0$ implies that the differences from the counting argument are just energies themselves, and so $k = r$ therefore yields $D_{\seq}^{\EP}(n) = \Theta(n^r).$ To compute the collective cost, we take $A = E$ and set $a_0$ to be any element of $E$, say $a_0 = \epsilon_0$. The energy differences are then of the form $\epsilon - \epsilon_0  = (\epsilon - \epsilon_0) - (\epsilon' - \epsilon_0)$, and so they generate all gaps. Consequently, $k = s$ and $|nE| = \Theta(n^s)$. Since $D_{\cmp}^{\EP}(n)$ exceeds $|nE|$ by at most one, we also have $D_{\cmp}^{\EP}(n) = \Theta(n^s)$.
\end{proof}

\emph{Lifetime and sharing advantage.} The sequential cost is non-decreasing in $n$, since $\Gamma_n\subseteq\Gamma_{n+1}$. Therefore, inverting $D_{\seq}^{\EP}(n) = \Theta(n^r)$ yields $N_d=\Theta(d^{1/r})$. Since $N_{\fresh}(d)=\lfloor\log_ad\rfloor$, the sharing advantage in Eq.~\eqref{eq:advantage} scales as $\mathcal A(d) = \Theta(d^{1/r}/\log d)$. 

\section{Minimal example of separation for noisy operations} 
\label{app:finite}

\begin{figure*}
  \includegraphics{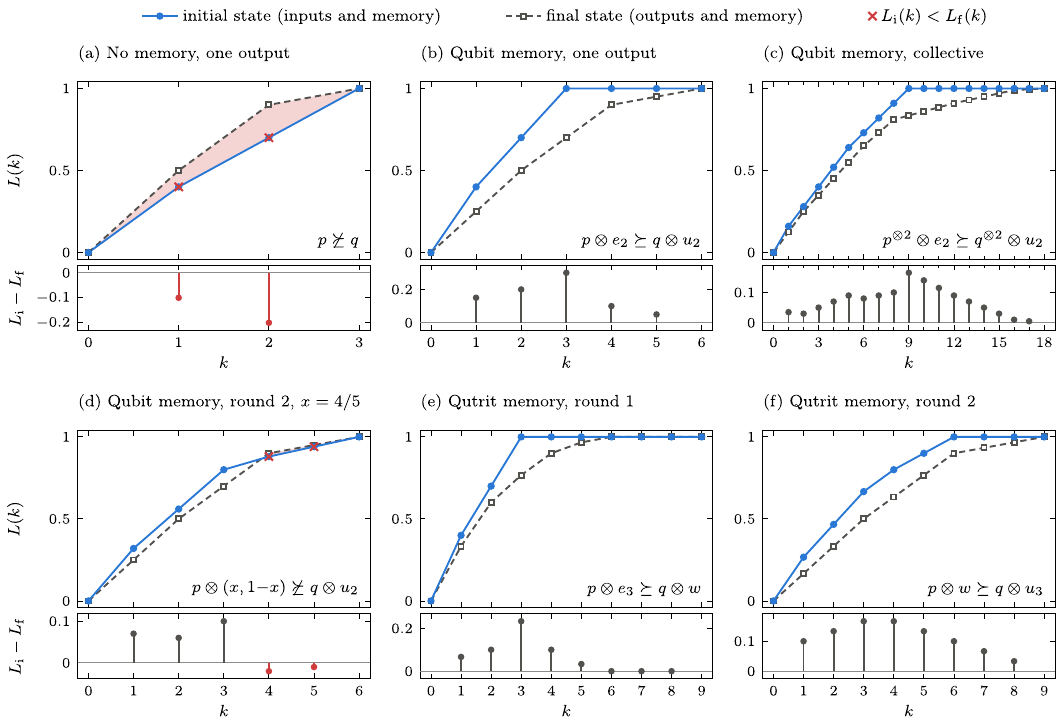}
  \caption{Lorenz curves for the minimal example studied in Appendix~\ref{app:finite}.   Each panel compares an initial state (inputs and memory) with a final   state (outputs and memory), where $L(k)$ is the sum of the $k$ largest   eigenvalues. A noisy operation can perform the conversion when $L_\mathrm{i}(k)\ge L_\mathrm{f}(k)$ for all $k$. The lower plots show $L_\mathrm{i}-L_\mathrm{f}$, with violations in red.   (a)~Without a memory, a single output cannot be prepared. A pure qubit memory suffices for (b)~one output and (c)~two outputs prepared collectively. (d)~Sequentially, the second round fails for   $x=4/5$, the largest memory eigenvalue allowed by the first round. (e,f)~A qutrit memory suffices for releasing both outputs.}
  \label{fig:lorenz}
\end{figure*}

Here we show Eq.~\eqref{eq:finitegap} for the spectra $p$ and $q$ of Eq.~\eqref{eq:finitepq}, for both output requirements, namely
\begin{align}
  p&=\Big(\tfrac25,\tfrac3{10},\tfrac3{10}\Big),
  &
  q&=\Big(\tfrac12,\tfrac25,\tfrac1{10}\Big).
\end{align}
We also demonstrate the two bounds on the intermediate memory state stated in Sec.~\ref{sec:finite}. For clarity, we drop the superscript $\NO$ throughout this section. The results of this elementary yet illustrative analysis is presented in Fig~\ref{fig:lorenz}.

We use standard characterization of noisy operations in terms of majorization~\cite{Gour2015}. More specifically, a noisy operation can map a state with spectrum $p$ to a state with spectrum $q$ if and only if $p$ majorizes\footnote{Here $v\succeq w$ means that, for every $k$, the sum of the $k$ largest entries of $v$ is at least the sum of the $k$ largest entries of $w$.} $q$, denoted $p\succeq q$. In particular, the entropy of a state cannot decrease under noisy operations. We further write $e_d=(1,0,\ldots,0)$ and $u_d=(1/d,\ldots,1/d)$ for the spectra of pure and maximally mixed $d$-level states respectively. To make each majorization easy to check, we list both spectra in decreasing order and rescale them to integers, with $c^{\times k}$ denoting an entry $c$ repeated $k$ times. We now compute all relevant memory costs.

\emph{Fresh cost.}
A single output requires a two-level memory. A one-dimensional memory fails, since $p_1<q_1$ implies $p\not\succeq q$. A two-level memory in a pure state suffices, since $p\otimes e_2\succeq q\otimes u_2$:
\begin{align}
  p\otimes e_2&=\frac{1}{20}(8,6,6,0,0,0),\notag\\
  q\otimes u_2&=\frac{1}{20}(5,5,4,4,1,1).
\end{align}
Hence $a=D_{\seq}(1)=2$ and $D_{\fresh}(2)=a^2=4$.

\emph{Collective cost.}
A two-level memory suffices to prepare two independent outputs, since $p^{\otimes2}\otimes e_2\succeq q^{\otimes2}\otimes u_2$. We can then write
\begin{align}
  p^{\otimes2}\otimes e_2
  &=\frac{1}{200}(32,24^{\times4},18^{\times4},0^{\times9}),\notag\\
  q^{\otimes2}\otimes u_2
  &=\frac{1}{200}(25^{\times2},20^{\times4},16^{\times2},5^{\times4},4^{\times4},1^{\times2}).
\end{align}
The final memory state is then maximally mixed and uncorrelated with the outputs. Without a memory, the two outputs would have entropy at least $2 H(p)$ bits, because noisy operations cannot decrease the entropy. By subadditivity, however, outputs with marginals $q$ have entropy at most $2H(q)$ bits. Hence $D_{\cmp}(2)=2$, even when only the marginals are required.

\emph{A two-level sequential memory fails.}
Suppose that a two-level memory prepares two outputs with the correct marginals, for some initial state $\omega_0$ and round-dependent noisy operations. Let $\tau$ be the joint state of the first output and the memory after the first round, and let $x\ge1/2$ be the largest eigenvalue of the memory state $\omega_1$ obtained at the end of the first round. We are going to show that the two rounds impose incompatible conditions on $x$. More specifically, the first round must leave the memory too mixed for the second round to be successful.

In the first round, the two largest eigenvalues of $\rho\otimes\omega_0$ sum to $p_1+p_2 \leq 0.7$, with equality for a pure $\omega_0$. By majorization, the same holds for $\tau$. Let $P$ project onto the eigenvector of $\omega_1$ with eigenvalue $x$, and let $Q$ project onto the eigenvectors of $\sigma$ with eigenvalues $q_1$ and $q_2$. Then we can write
\begin{align}
  x&=\tr[(I\otimes P)\tau]\notag\\
  &\le\tr[(Q\otimes P)\tau]+\tr[((I-Q)\otimes I)\tau]\notag\\
  &\le\tfrac7{10}+q_3=\tfrac45.\label{eq:firstbound}
\end{align}
The first term in the second line is the weight of $\tau$ on a two-dimensional subspace, and hence at most $0.7$. The second term is the probability $q_3=0.1$ that the output lies outside the range of $Q$.

The second round acts on $\rho\otimes\omega_1$, irrespectively from the correlations with the first output. To simplify, note we can replace the final memory by a maximally mixed state, which is a noisy operation and yields $\sigma\otimes u_2$. Hence
\begin{align}
  p\otimes(x,1-x)\succeq q\otimes u_2=\tfrac1{20}(5,5,4,4,1,1).
\end{align}
The largest entry on the left is $\tfrac25x$, so $x\ge\tfrac58$. For such $x$, the four largest entries on the left are $\tfrac25x$, $\tfrac3{10}x$, $\tfrac3{10}x$, and $\tfrac25(1-x)$. Their sum must be at least $\tfrac{18}{20}$, which gives $x\ge\tfrac56$. This contradicts Eq.~\eqref{eq:firstbound}, so $D_{\seq}(2)\ge3$ for both output requirements.

\emph{A three-level sequential memory suffices.}
Let the memory start in a pure state and pass through an intermediate state with spectrum $w=(2/3,1/3,0)$. Both rounds are then possible, since $p\otimes e_3\succeq q\otimes w$ and $p\otimes w\succeq q\otimes u_3$,
\begin{align}
  p\otimes e_3&=\frac{1}{30}(12,9,9,0^{\times6}),\notag\\
  q\otimes w&=\frac{1}{30}(10,8,5,4,2,1,0^{\times3}),\notag\\
  p\otimes w&=\frac{1}{30}(8,6,6,4,3,3,0^{\times3}),\notag\\
  q\otimes u_3&=\frac{1}{30}(5^{\times3},4^{\times3},1^{\times3}).
\end{align}
For respective plots in terms of Lorenz curves see Fig.~\ref{fig:lorenz}. Each round leaves its output uncorrelated with the memory, so the two outputs are independent. Hence $D_{\seq}(2)=3$ for both output requirements. Together with the fresh and collective costs, this proves Eq.~\eqref{eq:finitegap}, with $\Delta(2)=\log_2(3/2)$.

\section{Proof of Theorem~\ref{thm:family}}
\label{app:family}
In this Appendix we prove Theorem~\ref{thm:family} and Corollary~\ref{cor:qutrit}, and derive the sequential bound from Eq.~\eqref{eq:familyupper}. In what follows we denote the spectra of $\rho$ and $\sigma$ with $p$ and $q$ respectively. We again use $e_2=(1,0)$ and $u_2=(1/2,1/2)$ for the spectra of a pure and a maximally mixed qubit. Since both states have full rank, $p$ and $q$ are strictly positive
probability vectors of length $b$. Throughout this appendix we also denote $(a)_+:=\max\{a,0\}$

\subsection{Reduction to a majorization statement}
Theorem~\ref{thm:family} follows from a majorization argument that is formalized as Lemma~\ref{lem:familymaj} below. We prove the lemma in
Appendix~\ref{app:family-proof}.

\begin{lemma}\label{lem:familymaj}
If $p$ and $q$ satisfy condition~(i) of Theorem~\ref{thm:family}, then
there is an $n_0$ such that
\begin{align}
  p^{\otimes n}\otimes e_2\succeq q^{\otimes n}\otimes u_2
  \qquad\text{for every }n\ge n_0.
  \label{eq:familymajorization}
\end{align}
\end{lemma}

\begin{proof}[Proof of Theorem~\ref{thm:family}] The state $\rho^{\otimes n}\otimes\proj{0}$ has spectrum $p^{\otimes n}\otimes e_2$, and the state $\sigma^{\otimes n}\otimes u_2$, where $u_2$ is the maximally mixed qubit, has spectrum $q^{\otimes n}\otimes u_2$. A noisy operation can map one state to another of the same dimension whenever the spectrum of the first majorizes that of the second~\cite{Gour2015}. Hence, Lemma~\ref{lem:familymaj} guarantees the existence of a noisy operation implementing
\begin{align}
  \rho^{\otimes n}\otimes\proj{0}
  \longmapsto
  \sigma^{\otimes n}\otimes u_2
  \label{eq:collectivefamily}
\end{align}
for every $n\ge n_0$. The outputs are then independent and uncorrelated with the memory, so $D_{\cmp}^{\iid,\NO}(n)\le2$. Without a memory, a noisy operation would have to map $\rho^{\otimes n}$ to $\sigma^{\otimes n}$. Such an operation is unital, since it acts between spaces of equal dimension, and therefore cannot decrease the smallest eigenvalue. It would thus require $\lambda_{\min}(\rho)^n\le\lambda_{\min}(\sigma)^n$, which contradicts condition~(ii). Hence $D_{\cmp}^{\iid,\NO}(n)=2$ for $n\ge n_0$. Finally, condition~(ii) and Corollary~\ref{thm:support} give $D_{\seq}^{\iid,\NO}(n)\to\infty$.
\end{proof}

\subsection{Majorization via thresholds for multi-copy states}
\label{app:family-thresholds}
To prove Eq.~\eqref{eq:familymajorization}, we first rewrite majorization as a family of inequalities, one for each value of a threshold $t\ge0$. After deriving exponential estimates for both sides, we verify these inequalities separately in three ranges of $t$, each of which uses a different part of condition~(i).

Let $v$ and $v'$ be vectors with nonnegative entries, equal lengths, and equal sums, and let $L_v(k)$ be the sum of the $k$ largest entries of $v$, i.e., the Lorenz curve of $v$. By definition, $v\succeq v'$ means that $L_v(k)\ge L_{v'}(k)$ for every $k$. For a threshold $t\ge0$, the total amount by which the entries of $v$ exceed $t$ is then
\begin{align}
  \sum_j(v_j-t)_+=\max_k\,[L_v(k)-kt].
\end{align}
Conversely, $L_v(k)=\min_{t\ge0}\{\sum_j(v_j-t)_++kt\}$, where the minimum is attained when $t$ equals the $k$th largest entry of $v$. Therefore, $v\succeq v'$ holds if and only if
\begin{align}
  \sum_j(v_j-t)_+\ge\sum_j(v'_j-t)_+
  \qquad\text{for every }t\ge0.
  \label{eq:family-criterion}
\end{align}
Geometrically, $t$ is a slope of the Lorenz curve $L_v$, and $\sum_j(v_j-t)_+$ is the intercept of the tangent to $L_v$ with that slope. Comparing Lorenz curves at every $k$ is thus equivalent to comparing their tangents at every slope $t$.

\begin{figure}[t]
  \includegraphics{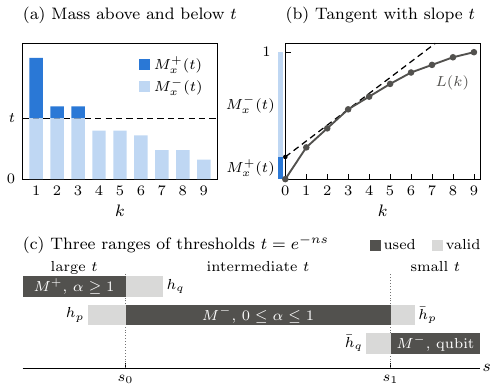}
  \caption{Quantities used in the proof of Lemma~\ref{lem:familymaj}.   (a)~Entries $x_w$ of $x^{\otimes2}$ for $x=(0.5,0.3,0.2)$, in decreasing order. The parts of the entries above and below the threshold $t$ add up to $M_x^+(t)$ and $M_x^-(t)$. (b)~Lorenz curve $L(k)$ of $x^{\otimes2}$. The tangent with slope $t$ touches the curve where $k$ equals the number of entries above $t$. It meets the vertical axis at height $M_x^+(t)$, a distance $M_x^-(t)$ below the final value $1$. (c)~Values of $s=-\frac1n\ln t$ for which each argument of Appendix~\ref{app:family-proof} is valid (light and dark) and for which it is used (dark), to scale for the example of Eq.~\eqref{eq:family}. Large thresholds are treated by comparing $M_p^+(t)$ with $M_q^+(2t)$ using the R\'enyi orders $\alpha\ge1$, intermediate thresholds by comparing $M_p^-(t)$ with $M_q^-(2t)$ using $0\le\alpha\le1$, and small thresholds with the auxiliary qubit. By condition~(i), consecutive arguments overlap on $(h_p,h_q)$ and on $(\bar h_q,\bar h_p)$, which contain the switching points $s_0$ and
  $s_1$.}
  \label{fig:thresholds}
\end{figure}

For a probability vector $x$ and a word $w=(w_1,\ldots,w_n)$, we write $x_w:=x_{w_1}\cdots x_{w_n}$ for the corresponding entry of $x^{\otimes n}$, and define
\begin{align}
  M_x^+(t)&:=\sum_w(x_w-t)_+,\notag\\
  M_x^-(t)&:=\sum_w\min\{x_w,t\},
  \label{eq:family-masses}
\end{align}
so that $M_x^+(t)+M_x^-(t)=1$. Intuitively, if each entry $x_w$ is drawn as a bar of height $x_w$, a horizontal cut at height $t$ divides the total probability into the part above the cut, $M_x^+(t)$, and the part below it, $M_x^-(t)$ [Fig.~\ref{fig:thresholds}(a)]. In terms of the Lorenz curve of $x^{\otimes n}$, the tangent with slope $t$ meets the vertical axis at height $M_x^+(t)$, which lies $M_x^-(t)$ below the final value $1$ of the curve [Fig.~\ref{fig:thresholds}(b)]. Raising the cut moves probability from above the cut to below it, so $M_x^+$ is nonincreasing and $M_x^-$ is nondecreasing in $t$.

We now apply Eq.~\eqref{eq:family-criterion} to the two sides of Eq.~\eqref{eq:familymajorization}. On the left-hand side, the entries are the $p_w$ together with $b^n$ zeros from $e_2$. Since $(0-t)_+=0$, the sum equals $M_p^+(t)$. On the right-hand side, every entry $q_w$ appears as two entries $q_w/2$, so the sum equals $2\sum_w(q_w/2-t)_+=M_q^+(2t)$. Hence,
Eq.~\eqref{eq:familymajorization} is equivalent to
\begin{align}
  M_p^+(t)\ge M_q^+(2t)\qquad\text{for every }t>0,
  \label{eq:family-threshold}
\end{align}
or, equivalently, to $M_p^-(t)\le M_q^-(2t)$ for every $t>0$.

The entries of $x^{\otimes n}$ decay exponentially with $n$, so we parametrize thresholds as $t=e^{-ns}$ with $s\in\mathbb R$. Larger values
of $s$ then correspond to smaller thresholds. Since $x_w=x_{w_1}\cdots x_{w_n}$, we can write
\begin{align}
  x_w>e^{-ns}
  \quad\Longleftrightarrow\quad
  \frac1n\sum_{i=1}^n\bigl(-\ln x_{w_i}\bigr)<s.
  \label{eq:family-average}
\end{align}
Thus, an entry exceeds the threshold exactly when the average of $-\ln x_{w_1},\ldots,-\ln x_{w_n}$ is smaller than $s$. To see which entries exceed the threshold, we compare $s$ with three statistics of the numbers $-\ln x_1,\ldots,-\ln x_b$, namely their minimum, their mean under $x$, and their mean under the uniform distribution:
\begin{align}
  z_x&:=-\ln\max_jx_j,\notag\\
  h_x&:=-\sum_jx_j\ln x_j,\notag\\
  \bar h_x&:=-\frac1b\sum_j\ln x_j.
  \label{eq:family-statistics}
\end{align}
Here $h_x$ is the Shannon entropy of $x$. The average in Eq.~\eqref{eq:family-average} is at least $z_x$, so no entry exceeds the threshold if $s\le z_x$. By the law of large numbers, as $n\to\infty$, the average concentrates around $h_x$ if the letters of a word are drawn independently from $x$, and around $\bar h_x$ if they are drawn uniformly. Consequently, the total probability of the entries above the threshold tends to $1$ if $s>h_x$ and to $0$ if $s<h_x$. Likewise, the fraction of entries above the threshold tends to $1$ if $s>\bar h_x$ and to $0$ if $s<\bar h_x$. Finally, $z_x\le h_x$, since a minimum cannot exceed a mean, and Jensen's inequality for the logarithm gives $h_x\le\ln b\le\bar h_x$.

\subsection{Exponential estimates}
\label{app:family-estimates}
For $s\ne h_x$, one of the two quantities $M_x^\pm(e^{-ns})$ tends to zero exponentially in $n$. We therefore compare the two quantities through their exponential rates. For $\alpha\in\mathbb R$, let $K_x(\alpha):=\ln\sum_jx_j^\alpha$, which is convex in $\alpha$, and
define
\begin{align}
  g_x(\alpha,s):=K_x(\alpha)+(\alpha-1)s,
\end{align}
together with the two exponents
\begin{align}
  \mu_x^+(s)&:=\inf_{\alpha\ge1}g_x(\alpha,s),
  &
  \mu_x^-(s)&:=\min_{0\le\alpha\le1}g_x(\alpha,s).
  \label{eq:family-rates}
\end{align}
The following lemma shows that these exponents are the exponential rates of the two sums in Eq.~\eqref{eq:family-masses}. In particular, $e^{n\mu_x^+(s)}$ is, to leading exponential order, the total probability of the entries of $x^{\otimes n}$ above $e^{-ns}$.

\begin{lemma}\label{lem:family-estimates}
Let $x$ be a strictly positive probability vector of length $b$. For
every $n$ and $s$, we have
\begin{align}
  M_x^+(e^{-ns})&\le e^{n\mu_x^+(s)},\notag\\
  M_x^-(e^{-ns})&\le e^{n\mu_x^-(s)}.
  \label{eq:family-upper}
\end{align}
Moreover, for every $\varepsilon>0$ and every compact interval $I$, we  have the following bounds
\begin{align}
  M_x^+(e^{-ns})&\ge e^{n[\mu_x^+(s)-\varepsilon]},\notag\\
  M_x^-(e^{-ns})&\ge e^{n[\mu_x^-(s)-\varepsilon]},
  \label{eq:family-lower}
\end{align}
which hold for all sufficiently large $n$ and $s \in I$, with the first bound valid on $I\subset(z_x,\infty)$.
\end{lemma}

\begin{proof}
For $y,t>0$, we have $(y-t)_+\le y^\alpha t^{1-\alpha}$ if $\alpha\ge1$, and $\min\{y,t\}\le y^\alpha t^{1-\alpha}$ if $0\le\alpha\le1$. Summing over words with $t=e^{-ns}$ and using $\sum_wx_w^\alpha=e^{nK_x(\alpha)}$ gives $M_x^\pm(e^{-ns})\le e^{ng_x(\alpha,s)}$ for every $\alpha$ in the respective range. Optimizing over $\alpha$ leads to Eq.~\eqref{eq:family-upper}.
 
The lower bounds follow from Cram\'er's theorem~\cite{Dembo2010}. Suppose we draw the letters of a word $w$ independently, either from $x$ or uniformly, and denote the corresponding probabilities by $\mathrm{Pr}_x$ and $\mathrm{Pr}_u$. Then
\begin{align}
  M_x^+(t)&\ge\tfrac12\mathrm{Pr}_x[x_w\ge2t],\notag\\
  M_x^-(t)&=\mathrm{Pr}_x[x_w\le t]+b^nt\,\mathrm{Pr}_u[x_w>t].
\end{align}
Here $-\ln x_w$ is a sum of $n$ independent terms, with cumulant generating function $K_x(1-\lambda)$ under $\mathrm{Pr}_x$ and $K_x(-\lambda)-\ln b$ under $\mathrm{Pr}_u$. Cram\'er's theorem gives the exponential rates of these probabilities as Legendre transforms of these functions. For $s>z_x$, the rate of $\mathrm{Pr}_x[x_W\ge2t]$ is $\inf_{\alpha\ge1}g_x(\alpha,s)=\mu_x^+(s)$. The rates of the two terms
in $M_x^-(t)$ are $\inf_{\alpha\le1}g_x(\alpha,s)$ and $\inf_{\alpha\ge0}g_x(\alpha,s)$. Since $g_x$ is convex in $\alpha$, one of these infima is attained in $[0,1]$, so the larger of the two equals $\mu_x^-(s)$. Finally, $M_x^\pm$ are monotone in $t$ and the rates are continuous in $s$, so the convergence is uniform on compact intervals.
\end{proof}

\subsection{Proof of Lemma~\ref{lem:familymaj}}
\label{app:family-proof}
We start by investigating the consequences of condition~(i). Using our notation, for $\alpha\neq1$, the R\'enyi entropies can be written as $H_\alpha(x)=K_x(\alpha)/[(1-\alpha)\ln2]$. Condition~(i) then
states that
\begin{align}
  K_p(\alpha) &< K_q(\alpha)\quad\text{for }0<\alpha<1,\notag\\
  K_p(\alpha) &> K_q(\alpha)\quad\text{for }\alpha>1,
  \label{eq:family-conditions}
\end{align}
together with $h_p<h_q$, $z_p<z_q$, and $\bar h_p>\bar h_q$. These
three inequalities correspond to $\alpha=1$, to $\alpha=\infty$, and to
the determinant condition, respectively. Moreover, $g_x(0,s)=\ln b-s$ and $g_x(1,s)=0$, with slopes given by
\begin{align}
  \partial_\alpha g_x(0,s)&=s-\bar h_x,
  &
  \partial_\alpha g_x(1,s)&=s-h_x.
  \label{eq:family-slopes}
\end{align}

Let us now choose $s_0\in(h_p,h_q)$ and $s_1\in(\bar h_q,\bar h_p)$. Since $h_q\le\ln b\le\bar h_q$, we have $s_0<s_1$. These two values split the thresholds $t=e^{-ns}$ into three ranges [Fig.~\ref{fig:thresholds}(c)]. For $s\le s_0$, the total probability of the entries of $q^{\otimes n}$ above the threshold tends to $0$, since $s_0<h_q$. There, we compare $M_p^+$ with $M_q^+$ using the orders $\alpha\ge1$, including $\alpha=\infty$. For $s_0\le s\le s_1$, we compare $M_p^-$ with $M_q^-$ using the orders $\alpha\in[0,1]$. For $s\ge s_1$, the fraction of entries of $q^{\otimes n}$ above the threshold tends to $1$, since $s_1>\bar h_q$. There, we use the qubit. We call these thresholds large, intermediate, and small, respectively, and verify Eq.~\eqref{eq:family-threshold} for each of them.

\emph{Large thresholds $(s\le s_0)$.}
Since $M_q^+(2t)\le M_q^+(t)$, it suffices to show $M_p^+(t)\ge M_q^+(t)$. For $s\le z_q$, every entry satisfies $q_w\le e^{-nz_q}\le t$, so $M_q^+(t)=0$. For $s\in[z_q,s_0]$, we show that $\mu_p^+(s)>\mu_q^+(s)$. If $s\ge h_p$, Eq.~\eqref{eq:family-slopes} and convexity give $\mu_p^+(s)=g_p(1,s)=0$, whereas $\mu_q^+(s)<0$ because $s<h_q$. If $s<h_p$, the infimum defining $\mu_p^+(s)$ is attained at some $\alpha_p>1$, because $g_p(\alpha,s)\ge\alpha(s-z_p)-s\to\infty$ as $\alpha\to\infty$. Then \begin{align}
  \mu_q^+(s)\le g_q(\alpha_p,s)<g_p(\alpha_p,s)=\mu_p^+(s).
\end{align}
The difference $\mu_p^+-\mu_q^+$ is continuous on $(z_p,\infty)$, and $\mu_q^+$ is an infimum of affine functions of $s$. The difference therefore attains a minimum $\gamma>0$ on this compact interval. Since $[z_q,s_0]\subset(z_p,\infty)$, we can use Lemma~\ref{lem:family-estimates} with $\varepsilon=\gamma/2$ to obtain
\begin{align}
  M_q^+(t)\le e^{n\mu_q^+(s)}<e^{n[\mu_p^+(s)-\gamma/2]}\le M_p^+(t)
\end{align}
for all sufficiently large $n$.

\emph{Intermediate thresholds $(s_0\le s\le s_1)$.}
Since $M_q^-(2t)\ge M_q^-(t)$, it suffices to show $M_p^-(t)\le M_q^-(t)$. We show that $\mu_p^-(s)<\mu_q^-(s)$. Let $\alpha_q\in[0,1]$ attain the minimum defining $\mu_q^-(s)$. If $0<\alpha_q<1$, then we can write
\begin{align}
  \mu_p^-(s)\le g_p(\alpha_q,s)<g_q(\alpha_q,s)=\mu_q^-(s).
\end{align}
It remains to check the endpoints. If $\alpha_q=1$, then $\mu_q^-(s)=0=g_p(1,s)$, whereas $\mu_p^-(s)<g_p(1,s)$ because $\partial_\alpha g_p(1,s)=s-h_p>0$. If $\alpha_q=0$, then $\mu_q^-(s)=\ln b-s=g_p(0,s)$, and $\mu_p^-(s)<g_p(0,s)$ because $\partial_\alpha g_p(0,s)=s-\bar h_p<0$. Therefore $\mu_q^--\mu_p^-$ achieves minimum $\gamma>0$ on $[s_0,s_1]$. Lemma~\ref{lem:family-estimates}, with the choice $\varepsilon=\gamma/2$, then gives
\begin{align}
  M_p^-(t)\le e^{n\mu_p^-(s)}<e^{n[\mu_q^-(s)-\gamma/2]}\le M_q^-(t).
\end{align}
for all sufficiently large $n$, uniformly in $s$.

\emph{Small thresholds $(s\ge s_1)$.}
Here the exponential estimates are not enough to guarantee Eq.~\eqref{eq:family-threshold}. More specifically, for $s\ge\bar h_p$, both exponents $\mu_p^-(s)$ and $\mu_q^-(s)$ equal $\ln b-s$, so $M_p(-t)$ and $M_q(-2t)$ have the same exponential rate. The comparison is instead specified by the number of nonzero entries on each side of Eq.~\eqref{eq:familymajorization}. The pure qubit $e_2$ adds only zeros, so the left-hand side has $b^n$ nonzero entries. Each contributes at most $t$, and hence $M_p^-(t)\le b^nt$. The maximally mixed qubit $u_2$ splits every $q_w$ into two entries $q_w/2$, so the right-hand side has $2b^n$ nonzero entries. Every word with $q_w\ge2t$ contributes $\min\{q_w,2t\}=2t$ to $M_q^-(2t)$, so it suffices to show that more than half of all words satisfy $q_w\ge2t$. For that, consider words drawn uniformly at random. Then $-\frac1n\ln q_w$ is an average of $n$ independent terms with mean $\bar h_q<s_1$. By the law of large numbers, more than half of all $b^n$ words satisfy $q_w\ge e^{-n(\bar h_q+s_1)/2}$ for large $n$. For large $n$, this lower bound exceeds $2e^{-ns_1}\ge2t$. Hence
\begin{align}
  M_q^-(2t)>\tfrac12b^n\cdot2t=b^nt\ge M_p^-(t).
\end{align}

The three ranges (i.e. large, intermediate and small) cover all possible thresholds $t>0$, and each requires only that $n$ be sufficiently large. Therefore there exist $n_0$ such that in all three ranges Eq.~\eqref{eq:family-threshold} is satisfied. This therefore proves Lemma~\ref{lem:familymaj}. 

\subsection{Proof of Corollary~\ref{cor:qutrit}}
Observe that for qutrits $(b = 3)$, condition~(ii) of Theorem~\ref{thm:family} reads $p_3>q_3$. The $\alpha = \infty$ inequality from condition~(i) reads $p_1>q_1$. It remains to prove $H_\alpha(p)<H_\alpha(q)$ for $\alpha\in(0,\infty)$. For that, consider
\begin{align}
  f(\alpha):=\sum_{j=1}^3p_j^\alpha-\sum_{j=1}^3q_j^\alpha.
\end{align}
The required inequalities state that $f(\alpha)<0$ for $\alpha \in (0,1)$, $f(\alpha)>0$ for $\alpha \in (1,\infty)$, and additionally, for $\alpha=1$, that $f'(1)=h_q-h_p>0$.

Since $f$ is continuous, it can change sign only at its zeros, so locating all of them determines the sign of $f$ on $(0,1)$ and on $(1,\infty)$. In order to count the zeros of $f$, we proceed as follows. A function $\sum_{i=1}^mc_ia_i^\alpha$ with real coefficients $c_i$ and $a_1>a_2>\cdots>a_m>0$ has at most as many real zeros, counted with multiplicity, as the sequence $c_1,\ldots,c_m$ has sign changes. In $f$, each $p_j$ enters with coefficient $+1$ and each $q_j$ with $-1$. By the assumptions, $p_1>q_1\ge q_2\ge p_2\ge p_3>q_3$, so the coefficients in this order have the signs $+,-,-,+,+,-$, which change three times. If some of these numbers coincide, the corresponding terms merge, which cannot increase the number of sign changes. Hence $f$ has at most three zeros.

These zeros are easy to find. Two of them are $\alpha=0$ and $\alpha=1$, which is easy to verify. Moreover, $f'(0)=\ln(\det\rho/\det\sigma)<0$, so $f>0$ below $\alpha=0$. Since $q_3$ is the smallest of the six numbers, the term $-q_3^\alpha$ dominates as $\alpha\to-\infty$, so $f(\alpha)\to-\infty$. Hence $f$ has a third zero below $\alpha=0$. As $f'(0)<0$, $f$ is consequently negative on $(0,1)$. As $f$ changes sign at $\alpha=1$, it is positive on $(1,\infty)$, and $f'(1)>0$. This proves Corollary~\ref{cor:qutrit}.

\subsection{Sequential upper bound}
The block construction (discussed in detail in Appendix~\ref{app:rates}) turns the collective protocol of Eq.~\eqref{eq:collectivefamily} into a sequential one. Every block of $\ell\ge n_0$ outputs is prepared in $\sigma^{\otimes\ell}$ from fresh inputs and an unused memory qubit, so all released outputs are independent. Eq.~\eqref{eq:app-rates-block} with $z=2$ gives
\begin{align}
  \log_2D_{\seq}^{\iid,\NO}(n)
  &\le\ell\log_2b+\lceil n/\ell\rceil-1\notag\\
  &\le\ell\log_2b+n/\ell.
\end{align}

For large $n$, the choice $\ell=\lceil\sqrt{n/\log_2b}\,\rceil\ge n_0$
yields 
\begin{align}
  \log_2D_{\seq}^{\iid,\NO}(n)\le2\sqrt{n\log_2b}+\log_2b,
\end{align}
which proves Eq.~\eqref{eq:familyupper}.

\section{Proof of Theorem~\ref{thm:rates}}
\label{app:rates}
In what follows we write both costs for the same output requirement (i.e. either marginal or independent). To prove Theorem~\ref{thm:rates}, note that by Eq.~\eqref{eq:dseq-submultiplicative} and Fekete's lemma, both sequential and collective rates exist and are equal to their respective infima, that is
\begin{align}
  R_{\seq}&:=\lim_{n\to\infty}\frac{\log_2D_{\seq}(n)}{n}
  =\inf_{n\ge1}\frac{\log_2D_{\seq}(n)}{n},\notag\\
  R_{\cmp}&:=\lim_{n\to\infty}\frac{\log_2D_{\cmp}(n)}{n}
  =\inf_{n\ge1}\frac{\log_2D_{\cmp}(n)}{n}.
  \label{eq:app-rates-fekete}
\end{align}
It therefore remains to show that $R_{\seq}=R_{\cmp} = R$.

\begin{proof}[Proof of Theorem~\ref{thm:rates}] The proof consists of three parts. First, we connect collective to sequential preparation by bounding the sequential rate by the memory cost of the collective protocol. Then we optimize this bound over collective protocols, which gives $R_{\seq}\le R$. Finally, we compose the rounds of a sequential protocol into a collective one, which gives the reverse inequality $R\le R_{\seq}$.

\emph{Part I: From collective to sequential preparation.}
Fix a block size $\ell\ge1$ and a successful collective protocol for
$\ell$ outputs, with a channel $\Phi_\ell$ and a memory of dimension $z$
in the initial state $\nu$. We use it to construct a sequential protocol.
Its memory consists of a buffer $X=X_1\cdots X_\ell$, whose slots $X_j$
are physically identical to an input system and have dimension $b$, and
of $T\ge0$ registers $Z_1,\ldots,Z_T$, each a copy of the collective
memory. These are all the systems retained between rounds, so the memory
has dimension
\begin{align}
  \dim(XZ_1\cdots Z_T)=b^\ell z^T.
  \label{eq:app-rates-memory}
\end{align}
We initialize the buffer in $\sigma^{\otimes\ell}$ and every register
in $\nu$.

In every round, the protocol swaps the fresh input with the next slot of the buffer, which is allowed by (A2). This releases the content of the slot and stores the input in its place. During the first $\ell$ rounds, which we call cycle $0$, the stored targets are released one by one and the buffer collects $\ell$ fresh inputs. After these rounds, the buffer is in $\rho^{\otimes\ell}$.

The protocol then proceeds in cycles $t=1,\ldots,T$ of $\ell$ rounds each. At the start of cycle $t$, we apply $\Phi_\ell$ to the buffer and the unused register $Z_t$. We combine this map with the first swap of the cycle, so no additional round is needed. By (A1) and (A2), every round thus uses an allowed operation. Just before $\Phi_\ell$ acts, the buffer holds the $\ell$ inputs collected in cycle $t-1$, and $Z_t$ is still in $\nu$. Neither has interacted with the earlier outputs. After $\Phi_\ell$, the buffer is therefore in the state
\begin{align}
  \Omega_X=\tr_{Z_t}\Phi_\ell\bigl(\rho^{\otimes\ell}\otimes\nu\bigr),
\end{align}
which satisfies the chosen requirement and is uncorrelated with all earlier outputs. The swaps of cycle $t$ release $X_1,\ldots,X_\ell$ one at a time. Each swap acts only on one slot and a fresh input, so the remaining slots are unchanged until their release. The $\ell$ outputs of the cycle are therefore jointly in the state $\Omega_X$, independent of the outputs of all earlier cycles. Correlations between the slots, or between the slots and $Z_t$, thus never affect the released outputs. The used register $Z_t$ may remain correlated with these outputs, but it is never accessed again. After the cycle, the buffer is again in $\rho^{\otimes\ell}$, and the registers $Z_{t+1},\ldots,Z_T$ are still untouched.

The protocol thus releases $(T+1)\ell$ outputs satisfying the chosen requirement, using the memory in Eq.~\eqref{eq:app-rates-memory}. Since later rounds never act on earlier outputs, its first $n\le(T+1)\ell$ rounds form a successful protocol for $n$ outputs. Choosing $T=\lceil n/\ell\rceil-1$ gives, for every $n\ge1$,
\begin{align}
  \log_2D_{\seq}(n)
  \le\ell\log_2b+\bigl(\lceil n/\ell\rceil-1\bigr)\log_2z.
  \label{eq:app-rates-block}
\end{align}
The buffer contributes the fixed cost $\ell\log_2b$, which disappears from the rate as $n$ grows. Dividing by $n$ and letting $n\to\infty$ at fixed $\ell$, we obtain
\begin{align}
  R_{\seq}\le\frac{\log_2z}{\ell}.
  \label{eq:app-rates-concrete}
\end{align}
\emph{Part II: Optimization over collective protocols.}
For fixed $\ell$, choose an optimal collective protocol, so that
$z=D_{\cmp}(\ell)$. Equation~\eqref{eq:app-rates-concrete} then gives
\begin{align}
  R_{\seq}\le\frac{\log_2D_{\cmp}(\ell)}{\ell}.
\end{align}
Taking the infimum over $\ell$ yields $R_{\seq}\le R$. For this choice of
$z$, Eq.~\eqref{eq:app-rates-block} is the bound stated in Sec.~\ref{sec:rate}.

\emph{Part III: From sequential to collective preparation.}
Conversely, composing the round channels of an $n$-round sequential protocol, with identities on the inputs not yet used and on the outputs already released, gives a collective protocol. The sequential memory then serves as the collective memory, and therefore $D_{\cmp}(n)\le D_{\seq}(n)$, as in Eq.~\eqref{eq:prefix}. Dividing by $n$ and letting $n\to\infty$ gives $R\le R_{\seq}$.

Together with Part II, this proves $R_{\seq}=R_{\cmp}=R$, i.e., Eq.~\eqref{eq:rates}. Finally,
\begin{align}
  \frac{\Delta(n)}{n}
  =\frac{\log_2D_{\seq}(n)-\log_2D_{\cmp}(n)}{n}
  \longrightarrow0,
\end{align}
so that $\Delta(n)=o(n)$. This completes the proof of the theorem.
\end{proof}

\section{Reversible cooling}
\label{app:cooling}

We prove Eq.~\eqref{eq:coolrate} for full-rank states
$\rho$ and $\sigma$ on a $b$-dimensional system, with
$\Delta S:=S(\rho)-S(\sigma)>0$.

\emph{Lower bound.}
Consider a successful collective unitary protocol with a
$d$-dimensional memory initially in state $\nu$, and denote
its final joint state by $\Omega$. Unitary invariance and
subadditivity of entropy give
\begin{align*}
nS(\rho)+S(\nu)
&=S(\Omega_{B_1\cdots B_nC})\\
&\le \sum_{i=1}^n S(\Omega_{B_i})+S(\Omega_C)\\
&\le nS(\sigma)+\log_2 d.
\end{align*}
Since $S(\nu)\ge0$, we obtain
$D_{\cmp}^{\marg,\Uop}(n)\ge2^{n\Delta S}$.
The same bound holds for sequential preparation.

\emph{Achievability.}
Fix a block length $m$ and set
$q_m:=\floor{2^{m\Delta S}}+1$.
For a $q_m$-dimensional register $E$, define
\begin{align*}
x_m&:=\rho^{\otimes m}\otimes\proj{0}_E,\\
y_m&:=\sigma^{\otimes m}\otimes u_E,
\qquad u_E:=I_E/q_m.
\end{align*}
These states satisfy
\begin{align*}
S(y_m)-S(x_m)&=\log_2q_m-m\Delta S>0,\\
\rank x_m&=b^m\le b^mq_m=\rank y_m.
\end{align*}
The exact catalytic entropy result of
Ref.~\cite{Wilming2022,Wilming2021}, applied to the composite system
consisting of the block and $E$, therefore guarantees a
finite-dimensional catalyst $K_m$ in a state $\gamma_m$
and a unitary $V_m$ such that
\begin{align*}
\Theta_m&:=V_m(x_m\otimes\gamma_m)V_m^\dagger,\\
\tr_{K_m}\Theta_m&=y_m,\\
\tr_{B^mE}\Theta_m&=\gamma_m.
\end{align*}
Thus the entire block--register marginal is exactly $y_m$,
while correlations with $K_m$ are allowed.

Let $c_m:=\dim K_m$. To process $k$ blocks, use the memory
$C=K_mE_1\cdots E_k$, initially in
$\gamma_m\otimes(\proj{0})^{\otimes k}$.
Apply $V_m$ to each block, its register $E_j$, and the
same catalyst $K_m$. Before each application, the fresh
block and unused register are independent of $K_m$,
whose marginal remains $\gamma_m$. Hence every block
has the required output marginal, even though the
catalyst may be correlated with earlier outputs.
All registers $E_j$ are included in the initial memory
and retained after use. Consequently,
\begin{align*}
D_{\cmp}^{\marg,\Uop}(km)\le c_mq_m^k.
\end{align*}
Using the existence of the rate established in
Theorem~\ref{thm:rates}, and taking $k\to\infty$
at fixed $m$, gives
\begin{align*}
R_{\cmp}^{\marg,\Uop}
&\le \lim_{k\to\infty}
\frac{\log_2c_m+k\log_2q_m}{km}\\
&=\frac{\log_2q_m}{m}.
\end{align*}
Since
$\Delta S<\log_2q_m/m\le\Delta S+1/m$,
letting $m\to\infty$ yields
$R_{\cmp}^{\marg,\Uop}\le\Delta S$.
Together with the lower bound and
Theorem~\ref{thm:rates}, this proves
\begin{align*}
R_{\seq}^{\marg,\Uop}
=R_{\cmp}^{\marg,\Uop}
=\Delta S.
\end{align*}
The construction allows correlations between different
output blocks and therefore establishes the marginal
preparation rate.

Finally, the classical version of the same catalytic
result, Theorem~2 of Ref.~\cite{Wilming2022}, allows the
block conversion to use a diagonal auxiliary state and
a permutation in the input and output eigenbases.
The buffer-and-swap construction of
Theorem~\ref{thm:rates} preserves this classical
structure. Thus classical reversible protocols already
attain the optimal rate.

\section{Memory rate under noisy operations}
\label{app:noisy}

We prove Eq.~\eqref{eq:noisyrate} for full-rank states
$\rho$ and $\sigma$ on a $b$-dimensional system, with
$b\geq 2$. Let $p$ and $q$ denote their spectra, and set
\begin{equation}
R_* :=
\sup_{\alpha\in[0,\infty]}
\bigl[H_\alpha(p)-H_\alpha(q)\bigr]_+.
\end{equation}
By Theorem~\ref{thm:rates}, it suffices to determine the
collective rate for independent outputs.

\begin{proof}
\emph{Finite-block standard form.}
Write $e_d=(1,0,\ldots,0)$ and
$u_d=(1/d,\ldots,1/d)$ for the spectra of pure and
maximally mixed $d$-dimensional states. We claim that
\begin{equation}
D_{\cmp}^{\iid,\NO}(n)
=
\min\bigl\{
d\geq 1:
p^{\otimes n}\otimes e_d
\succeq
q^{\otimes n}\otimes u_d
\bigr\},
\label{eq:noisy_standard_form}
\end{equation}
where the minimum is over integers and $\succeq$ denotes
majorization.

Indeed, an independent-output protocol may leave the
outputs correlated with the memory. Completely
depolarizing the final memory is a noisy operation and
replaces the joint state by
$\sigma^{\otimes n}\otimes I_C/d$.
If the initial memory has spectrum $r$, the majorization
criterion for noisy operations~\cite{Gour2015} therefore gives
\[
p^{\otimes n}\otimes r
\succeq q^{\otimes n}\otimes u_d.
\]
Since $e_d\succeq r$, tensor compatibility and transitivity
imply the condition in Eq.~\eqref{eq:noisy_standard_form}.
Conversely, that condition yields a noisy operation with
a pure initial memory and the required product final state.

\emph{Lower bound.}
The nonnegative-order R\'enyi entropies are additive and
Schur concave, with $H_\alpha(e_d)=0$ and
$H_\alpha(u_d)=\log_2 d$. Hence
Eq.~\eqref{eq:noisy_standard_form} implies
\[
\log_2 D_{\cmp}^{\iid,\NO}(n)
\geq n\bigl[H_\alpha(p)-H_\alpha(q)\bigr]
\]
for every $\alpha\in[0,\infty]$. Taking the supremum
and using $D_{\cmp}^{\iid,\NO}(n)\geq 1$ gives
\[
\log_2 D_{\cmp}^{\iid,\NO}(n)\geq nR_*,
\]
and thus $R_{\cmp}^{\iid,\NO}\geq R_*$.

\emph{Upper bound.}
Choose positive integers $k,m$ such that $k/m>R_*$, and
define
\[
x=p^{\otimes m}\otimes e_{2^k},
\qquad
y=q^{\otimes m}\otimes u_{2^k}.
\]
For every $\alpha\in[0,\infty]$,
\begin{align}
H_\alpha(y)-H_\alpha(x)
&= k-m\bigl[H_\alpha(p)-H_\alpha(q)\bigr]
\nonumber\\
&\geq k-mR_*>0.
\end{align}
Since $p$ has full support and $b\geq 2$, we have
$0<H_\alpha(x)\leq m\log_2 b$. Consequently,
\begin{equation}
\inf_{\alpha\in[0,\infty]}
\frac{H_\alpha(y)}{H_\alpha(x)}
\geq
1+\frac{k-mR_*}{m\log_2 b}
>1.
\end{equation}
Jensen's sufficient condition for exact asymptotic
majorization~\cite{Jensen2019} therefore implies
$x^{\otimes t}\succeq y^{\otimes t}$ for sufficiently
large integers $t$; zero entries may be omitted when
applying this result. Regrouping the tensor factors gives
\[
p^{\otimes mt}\otimes e_{2^{kt}}
\succeq
q^{\otimes mt}\otimes u_{2^{kt}},
\]
so $D_{\cmp}^{\iid,\NO}(mt)\leq 2^{kt}$.
The infimum characterization in Theorem~\ref{thm:rates}
then yields
\[
R_{\cmp}^{\iid,\NO}
\leq
\frac{\log_2 D_{\cmp}^{\iid,\NO}(mt)}{mt}
\leq \frac{k}{m}.
\]
Letting $k/m$ approach $R_*$ from above proves the
matching upper bound, including when $R_*=0$.
Finally, Theorem~\ref{thm:rates} gives
\[
R_{\seq}^{\iid,\NO}
=
R_{\cmp}^{\iid,\NO}
=
R_*,
\]
which is Eq.~\eqref{eq:noisyrate}.
\end{proof}

\section{Proof of Theorem \ref{thm:stationary}}
\label{app:catalysis}
In this Appendix we prove Theorem~\ref{thm:stationary}. Throughout, we fix a stationary interaction $\mathcal E$ with memory and output channels $\Lambda$ and $\Sigma$, as defined in Eq.~\eqref{eq:reduced-channels}. Starting from $\omega_0$, the memory after $j$ rounds is in the state $\omega_j=\Lambda^j(\omega_0)$, and round $j+1$ releases an output with marginal $\Sigma(\omega_j)$. We say that this round succeeds if $\Sigma(\omega_j)=\sigma$.

The proof rests on two simple observations. First, success is a linear condition on the memory state, so it suffices to check it on a basis of the states that the memory visits. Second, the average of the memory state over many rounds is almost unchanged by one further round. Formalizing and combining these two observations will prove prove the theorem.

\begin{proof}[Proof of Theorem~\ref{thm:stationary}] We show (i)$\Rightarrow$(ii)$\Rightarrow$(iii)$\Rightarrow$(i).

\emph{(i)$\Rightarrow$(ii): Finitely many rounds enable all rounds.} Let us define the linear map 
\begin{align}
  \mathcal D(X):=\Sigma(X)-\tr(X)\,\sigma.
  \label{eq:app-cat-deviation}
\end{align}
For a density operator, $\tr(X)=1$, so round $j+1$ succeeds exactly when $\mathcal D(\omega_j)=0$. The memory states belong to the real vector space of Hermitian operators on $C$, which has dimension $d^2$. At most $d^2$ of them are therefore linearly independent, and there is a smallest $k\le d^2$ such that $\omega_k$ is a linear combination of $\omega_0,\ldots,\omega_{k-1}$. Let $\mathcal V$ be the span of $\omega_0,\ldots,\omega_{k-1}$. The channel $\Lambda$ maps these states to $\omega_1,\ldots,\omega_k$, which all lie in $\mathcal V$, so $\Lambda$ maps $\mathcal V$ into itself. Consequently, every memory state $\omega_j=\Lambda^j(\omega_0)$ lies in $\mathcal V$. By (i), $\mathcal D$ vanishes on $\omega_0,\ldots,\omega_{k-1}$, and hence, by linearity, on all of $\mathcal V$. Thus $\mathcal D(\omega_j)=0$ for every $j\ge0$, so every round succeeds for the same initial state, which is (ii). In other words, the memory can explore at most $d^2$ linearly independent directions, and every later state is a linear combination of states for which the preparation already succeeded.

\emph{(ii)$\Rightarrow$(iii): Averaging gives a catalyst.}
Suppose that the protocol succeeds in every round. Consider the average memory state over the first $N$ rounds, 
\begin{align}
  \bar\omega_N:=\frac1N\sum_{j=0}^{N-1}\omega_j.
\end{align}
By linearity we have $\Sigma(\bar\omega_N)=\sigma$. Moreover, one further round changes the average only by a small amount, 
\begin{align}
  \Lambda(\bar\omega_N)= \bar\omega_N + \frac{\omega_N-\omega_0}{N},
  \label{eq:app-cat-shift}
\end{align}
so that the trace norm between $\Lambda(\bar\omega_N)$ and $\bar\omega_N$ is at most $2/N$. The set of states on $C$ is compact, so a subsequence of the averages converges to a state $\omega$. By continuity of $\Lambda$ and $\Sigma$, we obtain $\Lambda(\omega)=\omega$ and $\Sigma(\omega)=\sigma$, so $\omega$ is a correlated catalyst. Note that the memory states $\omega_j$ themselves need not converge.

\emph{(iii)$\Rightarrow$(i): A catalyst has infinite lifetime.} If the memory starts in a correlated catalyst $\omega$, its reduced
state remains $\omega$ in every round, and each fresh input is independent of it. Every round therefore releases an output with marginal $\Sigma(\omega)=\sigma$. In particular, the first $d^2$ rounds
succeed.

\emph{Extension to uncorrelated catalysts.} The above arguments can be easily extended to the case when each round prepares the output uncorrelated with the memory. To see this, suppose that the first $d^2$ rounds release outputs uncorrelated with the memory, i.e.,
$\mathcal E(\rho\otimes\omega_j)=\sigma\otimes\omega_{j+1}$ for $j=0,\ldots,d^2-1$. This condition is again linear in the memory state. More specifically, it states that the linear map
\begin{align}
  \mathcal D'(X):=\mathcal E(\rho\otimes X)-\sigma\otimes\Lambda(X)
\end{align}
vanishes at $X=\omega_j$. As in the proof of (i)$\Rightarrow$(ii),
$\mathcal D'$ then vanishes on $\mathcal V$, and hence on every memory
state $\omega_j$. The proof then proceeds similarly as in the correlated case. Note that this situation requires each output to be uncorrelated with the current memory, which is stronger than requiring independent outputs. In a similar way one can prove that in this case (ii)$\Rightarrow$(iii) and (iii)$\Rightarrow$(i). 

\end{proof}
\end{document}